\pdfoutput=1
\documentclass{aa}  
\usepackage{amsmath}
\usepackage{graphicx}
\usepackage{txfonts}
\usepackage{natbib}
\bibpunct{(}{)}{;}{a}{}{,} 
\def\ie{{\it i.e.,}\,}
\def\eg{{\it e.g.,}\,}

\def\la{\hbox{\raise.5ex\hbox{$<$} 
    \kern-1.1em\lower.5ex\hbox{$\sim$}}} 
\def\ga{\hbox{\raise.5ex\hbox{$>$} 
    \kern-1.1em\lower.5ex\hbox{$\sim$}}}

\newcommand{\dgr}{\mbox{$^\circ$}}           
\newcommand{\Msun}{\mbox{$M_\odot$}}         
\newcommand{\Lsun}{\mbox{$L_\odot$}}         
\newcommand{\ad}{\mbox{d}}                   
\newcommand{\cm}{\mbox{\ cm}}                
\newcommand{\g}{\mbox{\ g}}                  
\newcommand{\s}{\mbox{\ s}}                  
\newcommand{\K}{\mbox{\ K}}                  
\newcommand{\erg}{\mbox{\ erg }}              
\newcommand{\cms}{\mbox{\ cm s${}^{-1}$}}    
\newcommand{\Ks}{\mbox{\ K s${}^{-1}$}}    
\newcommand{\Kcm}{\mbox{\ K cm${}^{-1}$}}    
\newcommand{\mes}{\mbox{\ m s${}^{-1}$}}    
\newcommand{\ergK}{\mbox{\ erg K${}^{-1}$}}    
\newcommand{\gcm}{\mbox{\ g cm${}^{-3}$}}    
\newcommand{\erggs}{\mbox{$\erg\g^{-1}\s^{-1}$}}  
\newcommand{\ergs}{\mbox{$\erg\s^{-1}$}}  
\newcommand{\ergcmsK}{\mbox{$\erg\K^{-1}\cm^{-1}\s^{-1}$}}   
\newcommand{\dyncm}{\mbox{\ dyn$\cm^{-2}$}}  

\newcommand{\sv}{\langle\sigma v\rangle}

\begin{document}
\bibliographystyle{aa}
   \title{The core helium flash revisited}

   \subtitle{I. One and two-dimensional hydrodynamic simulations}

   \author{M. Moc\'ak,
           E. M\"uller,
           A.Weiss,
           \and K.Kifonidis}


   \institute{Max-Planck-Institut f\"ur Astrophysik,
              Postfach 1312, 85741 Garching, Germany\\
              \email{mmocak@mpa-garching.mpg.de}}

   \date{Received  ........................... }

 
\abstract
   { We investigate the hydrodynamics of the core helium flash near
     its peak. Past research concerned with the dynamics of this event
     is inconclusive. However, the most recent multidimensional
     hydrodynamic studies suggest a quiescent behavior and seem to
     rule out an explosive scenario.}
   { Previous work indicated, that depending on initial conditions,
     employed turbulence models, grid resolution, and dimensionality
     of the simulation, the core helium flash leads either to the
     disruption of a low-mass star or to a quiescent quasi-hydrostatic
     evolution. We try to clarify this issue by simulating the
     evolution with advanced numerical methods and detailed
     microphysics.}
   { Assuming spherical or axial symmetry, we simulate the evolution
     of the helium core of a $1.25 M_{\odot}$ star with a metallicity
     Z=0.02 during the core helium flash at its peak with a grid-based
     hydrodynamics code.}
   { We find that the core helium flash neither rips the star apart,
     nor that it significantly alters its structure, as convection
     plays a crucial role in keeping the star in hydrostatic
     equilibrium. In addition, our simulations show the presence of
     overshooting, which implies new predictions concerning mixing of
     chemical species in red giants.}
   {}


   \maketitle

%

\section{Introduction}

In stars of mass 0.7 $\Msun$ $\le M \le$ 2.2 $\Msun$ the onset of
helium burning constitutes a major event -- the core helium flash. The
pre-flash stellar core contains a white dwarf-like degenerate
structure with a central density of about $10^6$\gcm, and an
off-center temperature maximum resulting from plasma- and
photo-neutrino cooling. When helium burning commences in this
degenerate core, the liberated nuclear energy cannot be used to expand
and cool the layers near the temperature maximum.  Instead it causes
further heating and a strong increase of the nuclear energy release.
Only when convection sets in, part of the excess energy can be
transported away from the burning regions, inhibiting thereby a
thermonuclear explosion. At the end of the flash, the core has been
expanded to densities of the order of $10^4$\gcm, with helium burning
quiescently in the center, and the star has settled on the horizontal
branch.  While standard stellar evolution calculations have been very
successful in reproducing observations of stars on the main sequence
and the red giant branch (RGB), we are forced to recognize several
discrepancies concerning the post-flash phases. In particular, we
recall the lack of understanding of the horizontal-branch morphology,
of low-luminosity carbon stars, and of hydrogen-deficient stars. Since
all these (and other) problems appear after the RGB phase, it is
plausible to suspect that the helium flash may be treated incorrectly
in standard (hydrostatic) stellar evolution calculations.

The conceptual problems associated with the helium core flash arise
from the extremely short timescales involved in the event. While the
pre-flash evolution proceeds on a nuclear timescale of $\sim$10$^8$
yrs, typical e-folding times for the energy release from helium
burning can become as short as hours at the peak of the flash. Such
short times are comparable to convective turnover times, \ie the
common assumptions used for the treatment of convection in stellar
evolution codes (instantaneous mixing, time-independence) are no
longer valid. In addition, the assumption of hydrostatic equilibrium
no longer needs to be fulfilled.  Early attempts to overcome these
assumptions by allowing for one-dimensional hydrodynamic flow
\citep{Edwards1969,Zimmermann1970, Villere1976,Wickett1977} remained
inconclusive. The results ranged from a confirmation of the standard
picture to a complete disruption of the star.

\citet{ColeDeupree1980,ColeDeupree1981} performed a two-dimensional
hydrodynamic study of the core helium flash. However, their study was
limited by the computational resources available at that time to a
rather coarse computational grid ($23\times4$ zones), a diffusive
first-order difference scheme (weighted donor cell), and a short time
evolution ($10^{5}$\,s compared to the duration of the core helium
flash of $~10^{11}$\,s\, from the onset of convection). They observed,
at the radius of the off-center temperature maximum, a series of
thermonuclear runaways where heat transport by convection and
conduction was sufficiently efficient to limit the rise of
temperature.  Each runaway modified the convective flow pattern and
led to some inward transport of heat across the off-center temperature
inversion. During the simulation the time interval between runaways
continuously shortened, and the maximum temperature steadily increased
until it eventually exceeded 10$^9K$.
 
\citet{DeupreeCole1983} and \citep{Deupree1984a,Deupree1984b}
confirmed these findings using two-dimensional models with an improved
angular resolution ($6\dgr$ instead of $20\dgr$), and
three-dimensional simulations (with $8\times8$ angular zones in a
$80\dgr \times 80\dgr$ cone, \ie $10\dgr$ angular resolution).
\citet{ColeDemDeupree1985} performed stellar evolution calculations of
the core helium flash using a model for convective overshooting based
on these hydrodynamic simulations. They found that the evolution of
the core helium flash is unchanged except for about the last week
prior to its peak. Furthermore, the possibility of mixing of core
material into the hydrogen shell was suggested by numerical
experiments where point source explosions were enforced
\citep{Deupree1984b,Deupree1986, DeupreeWallace1987}. These results
raised the hope that some problems concerning abundance anomalies and
mass loss could be solved by understanding the core helium flash.

The results of the hydrodynamic simulations, though varying in
details, indicated a dynamic flash that could disrupt the star
\citep{Deupree1984a} or at least lead to a significant loss of the
envelope \citep{ColeDeupree1981}. The simulations were critized by
\citet{IbenRenzini1984} and \citet{FujimotoIben1990} because (i) the
radial grid was too coarse, (ii) the gravitational potential was
``frozen in'' (\ie time-independent), and (iii) because a ``closed''
outer boundary was used. The latter two assumptions tend to
underestimate the expansion of the core, and hence tend to
overestimate the violence of the flash.

\begin{figure}
\includegraphics[width=0.99\hsize]{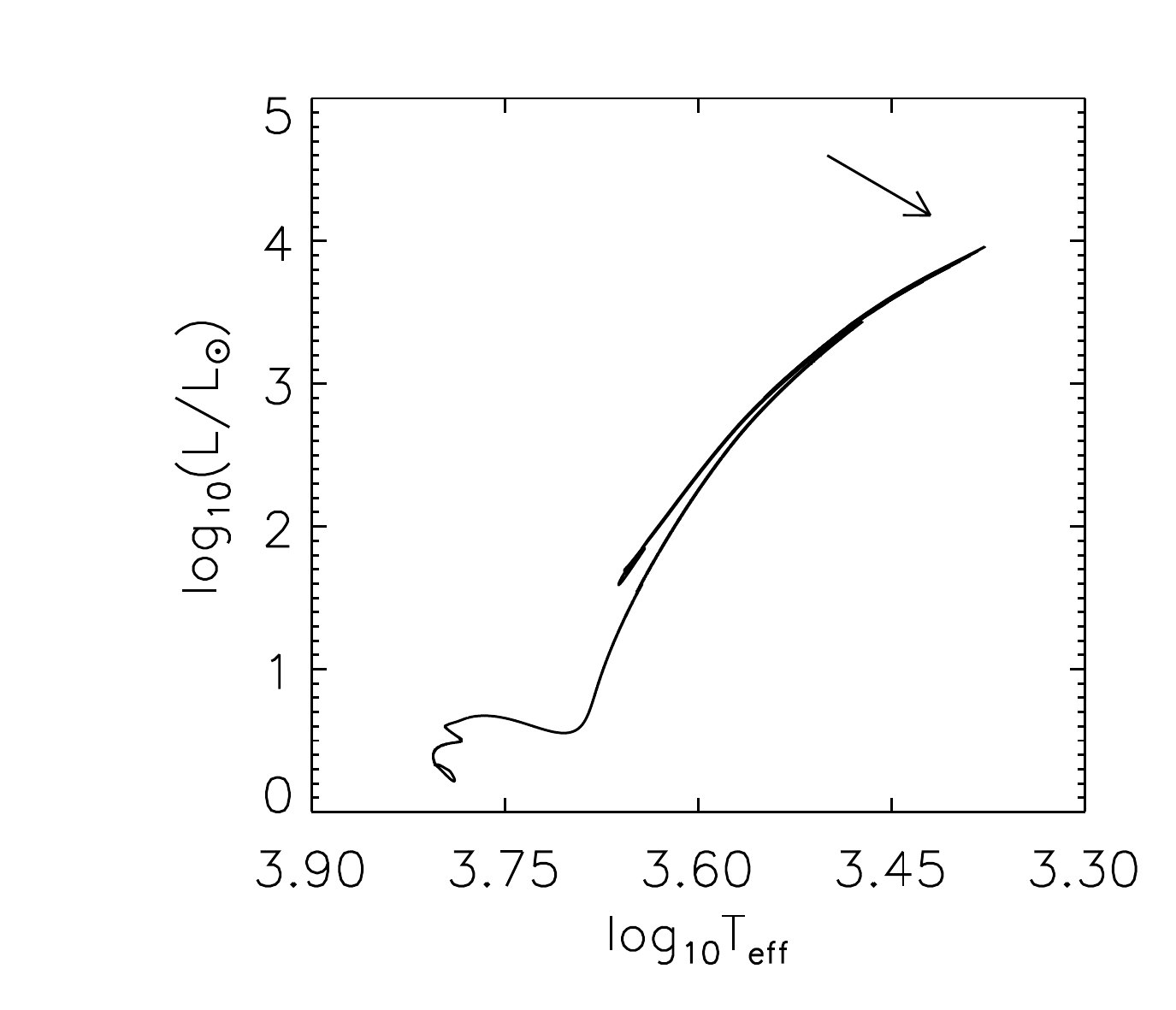}
\caption{Theoretical evolutionary track of a 1.25 $M_{\odot}$ star
         with a metallicity Z=0.02 in the H-R diagram. The core helium
         flash begins at the tip of the red giant branch indicated by
         the arrow.}
\label{fig1.1}
\end{figure}
 
Since the work of Deupree the computational capabilities have grown
tremendously and methods to simulate hydrodynamic flow have improved
considerably. Thus, the limitations of the early studies concerning
the grid resolution and the numerical treatment, which were the main
points of critique, meanwhile can be reduced considerably. At the same
time, we still have no coherent picture up to what extent and under
what circumstances (stellar mass and composition) hydrodynamic core
helium flash evolution could differ from canonical stellar evolution
calculations. It therefore appears necessary to have a new and fresh
look into the dynamics of the core helium flash. Incidentally,
\citet{Deupree1996} himself re-examined the problem already more than
a decade ago concluding that the flash does not lead to any
hydrodynamic event. Quiescent behavior of the core helium flash is
also favored by recent three-dimensional simulations
\citep{Dearborn2006, LattanzioDearborn2006} where the energy transport
due to convection, heat conduction, and radiation seems to be always
able to transport most of the energy generated during the flash
quiescently from the stellar interior to the outer stellar layers,
implying no hydrodynamic event, and hence a quasi-hydrostatic
evolution.
 
In the following we present a completely independent investigation of
the core helium flash by means of one-dimensional and two-dimensional
hydrodynamic simulations using state-of-the-art numerical techniques,
a detailed equation of state, and a time-dependent gravitational
potential.  The hydrodynamic calculations cover about 8 hrs of the
evolution near the peak of the core helium flash.  In passing we note
that the present investigation was instigated by a similar, meanwhile
technically obsolete study which was performed by Kurt Achatz 
\citep{Achatz1995} in the context of his diploma thesis. The results of this
latter study have unfortunately never been published.

The paper is organized as follows. In Sect.\,2 we discuss briefly the
stellar input model for the simulations along with some results from
hydrostatic core helium flash calculations.  In Sect.\,3 the
hydrodynamics code and the numerical methods are introduced, while the
results of our one and two-dimensional hydrodynamic runs are presented
in Sect.\,4 and 5, respectively. Finally, the conclusions are given in
Sect.\,6.


\section{Initial stellar models and hydrostatic calculations}

\begin{table*} 
\caption[]{Some properties of the initial model: total mass $M$,
           stellar population, metal content $Z$, mass $M_{He}$ and
           radius $R_{He}$ of the helium core ($X(^{4}He) > 0.98$),
           nuclear energy production in the helium core $L_{He}$,
           maximum temperature of the star $T_{max}$, and radius
           $r_{max}$ and density $\rho_{max}$ at the temperature
           maximum.}
\begin{tabular}{l|lllllllll} 
Model & $M$           & Pop.       & $Z$         & $M_{He}$     & $R_{He}$ 
      & $L_{He}$      & $T_{max}$  & $r_{max}$   & $\rho_{max}$ \\ 
      & $[\Msun]$     &            &             & $[\Msun]$    & $[10^9\cm]$ 
      & $[10^9\Lsun]$ & $[10^8\K]$ & $[10^8\cm]$ & $[10^5\gcm]$ \\
\hline 
M  & $1.25$ & I      & $0.02$ & $0.38$ & $1.91$ & $1.03$ 
   & $1.70$ & $4.71$ & $3.44$ \\

\end{tabular} 
\label{imodtab} 
\end{table*} 

\begin{figure} 
\includegraphics[width=0.99\hsize]{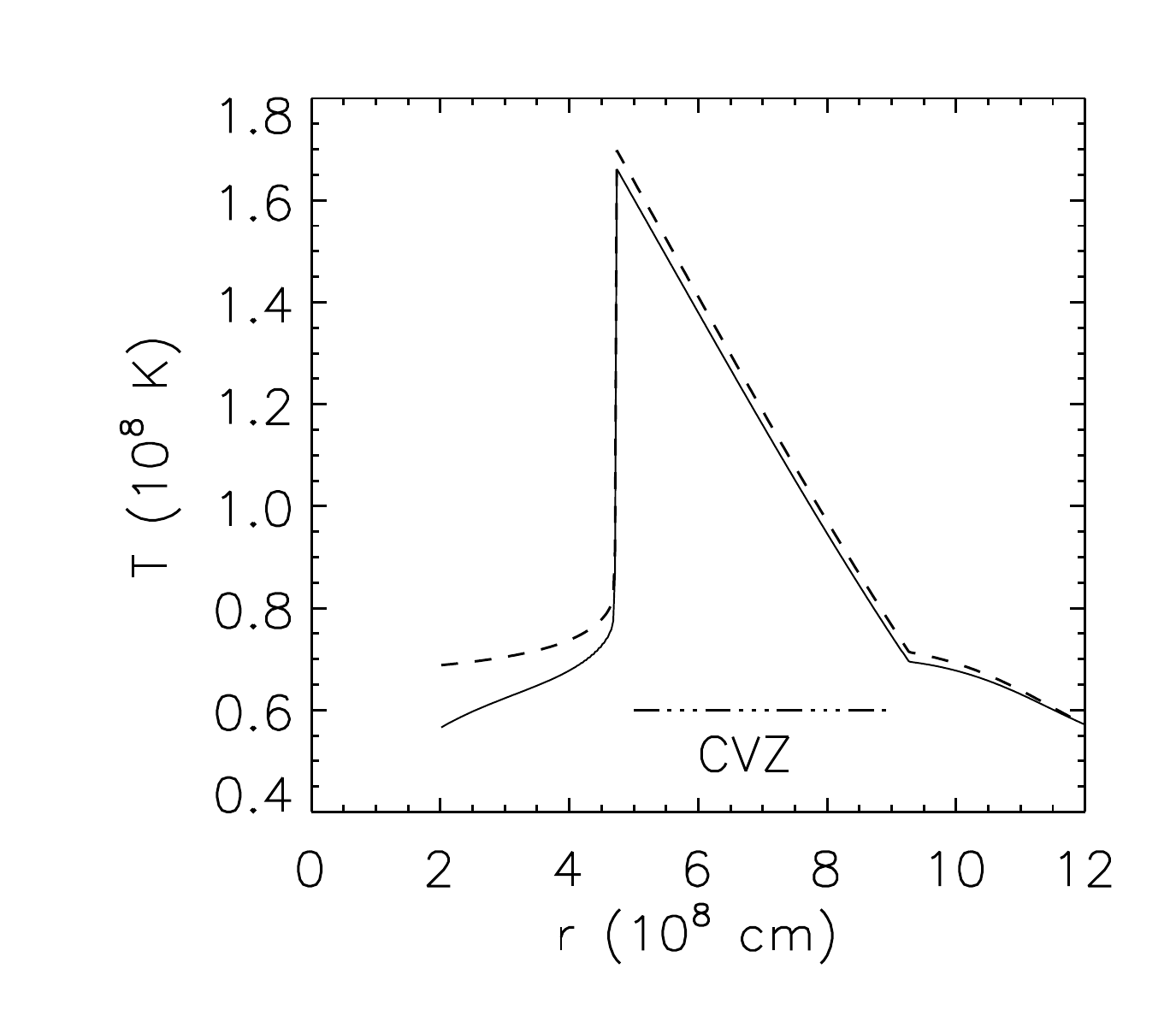}
\caption{Temperature distribution as a function of radius. The dashed
         line gives the distribution obtained from stellar
         evolutionary calculations with the ``Garstec'' code, while
         the solid line shows the mapped and stabilized distribution
         used as initial condition in the hydrodynamic
         simulations. CVZ marks the convection zone.}
\label{fig2.1}
\end{figure} 

\begin{figure*} 
\includegraphics[width=0.49\hsize]{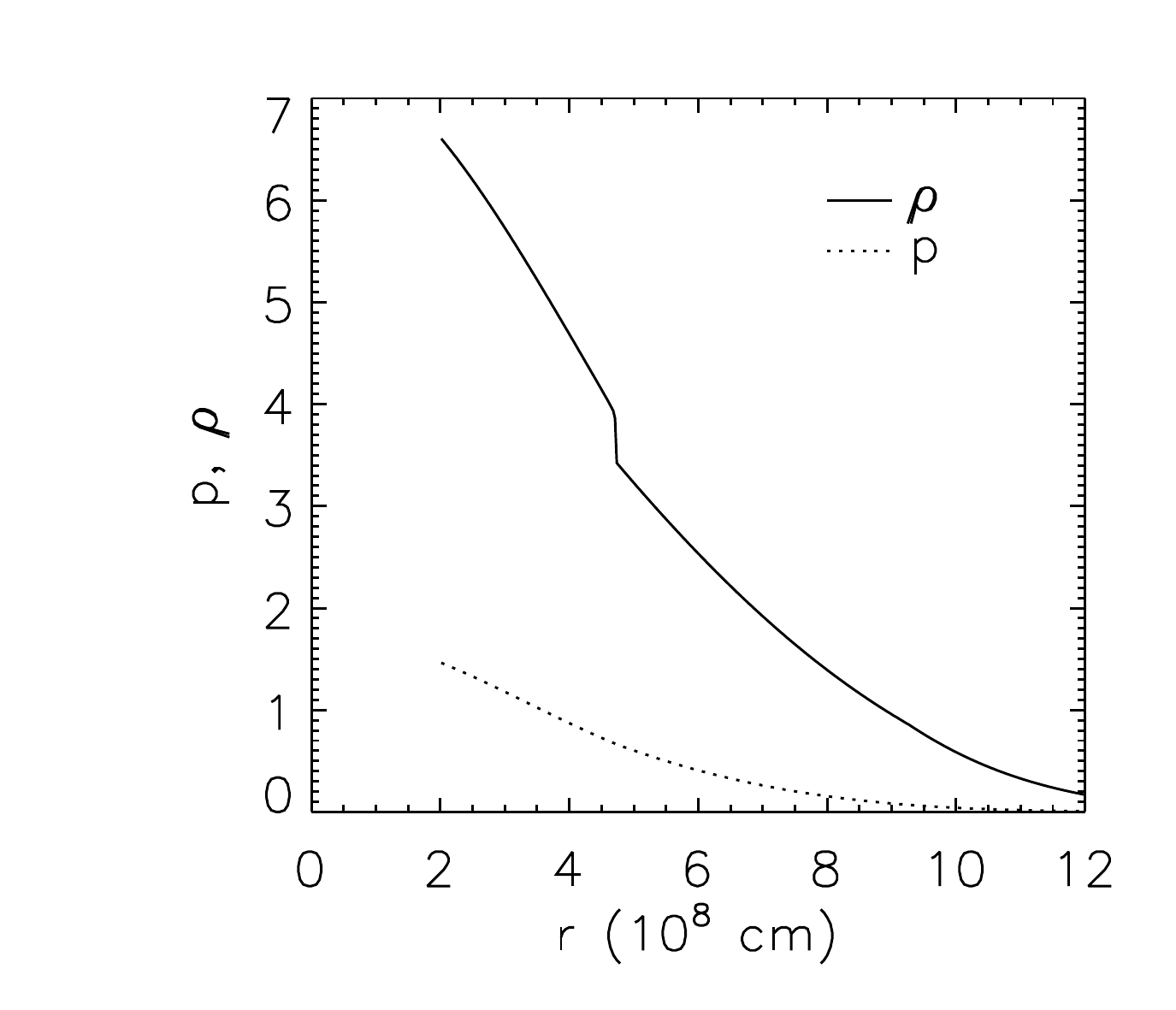} 
\includegraphics[width=0.49\hsize]{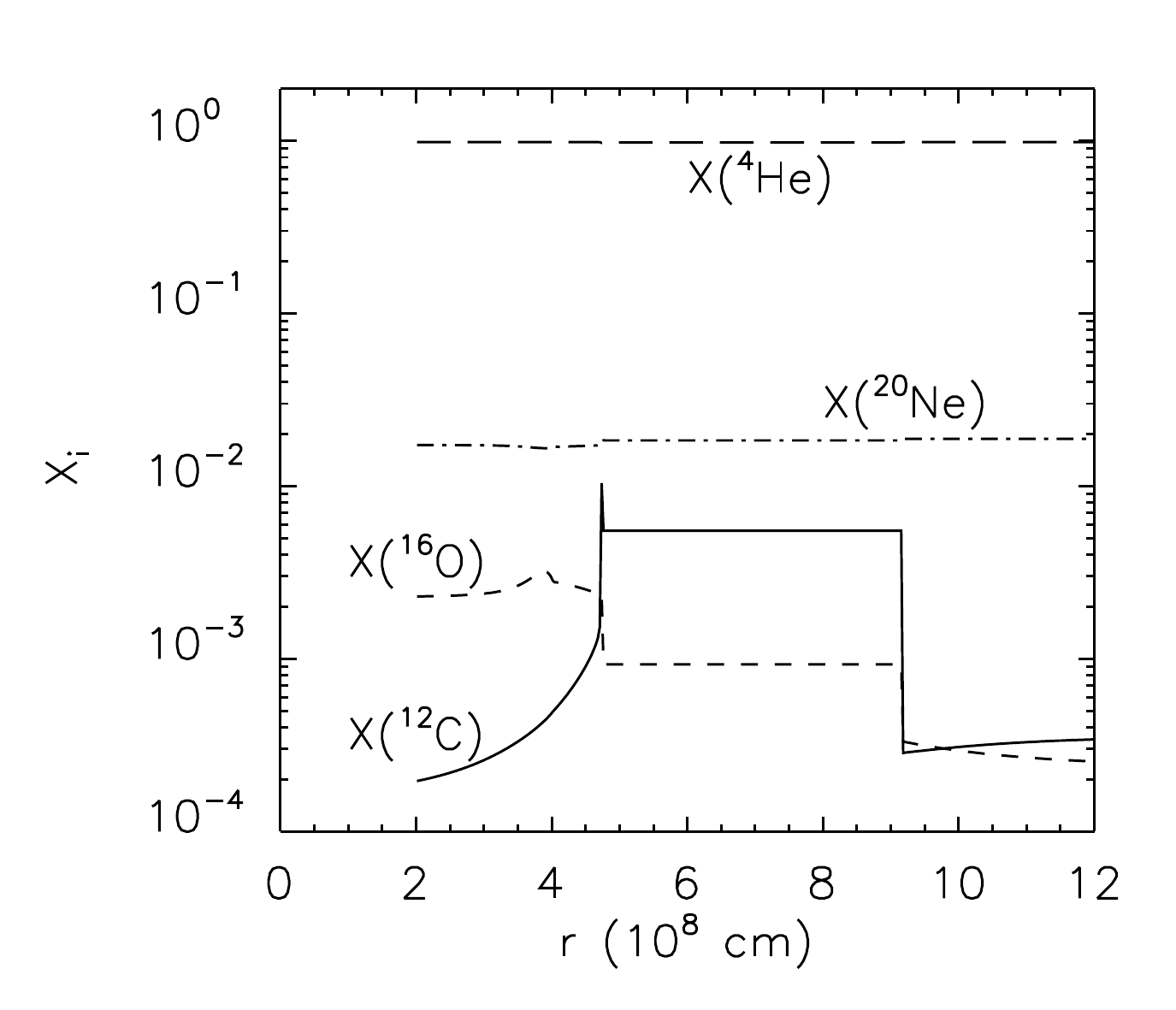} 
\caption{{\it{Left panel:}} Pressure (in $10^{22}\dyncm$) and density
         (in $10^5\gcm$) distribution of the mapped and stabilized
         initial model. The pressure and density profile of the
         original stellar evolution cannot be distinguished from the
         profiles of the mapped model on this scale.  {\it{Right
         panel:}} Chemical composition of the initial model showing a
         dominant fraction of helium and an apparent peak in $^{12}$C
         at the position of the temperature maximum resulting from a
         non-instantaneous treatment of convective mixing.}
\label{fig2.2.3}
\end{figure*} 

Table\,\ref{imodtab} summarizes some properties of our initial model,
which was obtained from stellar evolutionary calculations with the
``Garstec'' code \citep{WeissSchlattl2000, WeissSchlattl2007}.  It
corresponds to a star with a mass of 1.25 $\Msun$ and a metallicity Z
= 0.02 at the peak of the core helium flash ($L_{He} \sim\,
$10$^9\Lsun$) evolved with a hydrostatic stellar evolution code.
During this violent episode, the star is located at the tip of the red
giant branch in the H-R diagram (Fig.\,\ref{fig1.1}), hence being a
red giant consisting of a small central helium core with a radius
$r\sim\,$1.9$\, 10^{9}$ cm, surrounded by a hydrogen burning shell
and a huge convective envelope with a radius r$\,\sim 10^{13}$ cm.
Figure\,\ref{fig2.1} shows the temperature distribution inside the
helium core, which is characterized by an off-center temperature
maximum $T_{max}$, from where the temperature steeply drops towards
smaller radii and follows a super-adiabatic gradient towards larger
radii (convection zone). The radius $r_{max}$ of the temperature
maximum coincides with the bottom of the convection zone. The almost
discontinuous temperature stratification near $T_{max}$ (temperature
inversion), where the temperature rises from $7\,10^{7}$K to
$1.7\,10^{8}$K, results from an interplay between neutrino cooling and
heating by nuclear burning.  Figure\,\ref{fig2.2.3} shows the density
and pressure stratification of the model. One recognizes that the
temperature inversion is correlated with a drop in density. A detailed
view reveals that the steep increase of temperature corresponds to a
decrease of the density by 11\%, an increase of the ion pressure by
70\%, and a drop of the electron pressure by 9\%,
respectively.  Even at the peak of the core helium flash, the helium
core is still strongly degenerate: compared to the electron pressure
the ion pressure is lower by a factor of 6, while the radiation
pressure is smaller by almost 3 orders of magnitude.

The stellar model contains the chemical species $^{1}$H, $^{3}$He,
$^{4}$He, $^{12}$C, $^{13}$C, $^{14}$N, $^{15}$N, $^{16}$O ,$^{17}$O,
$^{24}$Mg, and $^{28}$Si.  However, since we are here not interested
in the detailed chemical evolution of the star, it is not necessary to
consider all of these species in our hydrodynamic simulations, as the
triple-$\alpha$ reaction dominates the energy production rate during
the core helium flash. For our hydrodynamic simulations we thus adopt
only the abundances of $^{4}$He, $^{12}$C, and $^{16}$O.  The
remaining composition is assumed to be adequately represented by a gas
with a mean molecular weight equal to that of $^{20}$Ne
(Fig.\,\ref{fig2.2.3}).

The stellar evolutionary model is one-dimensional, hydrostatic, and
was computed on a Lagrangian grid of 2294 zones. Since only the helium
core of the model (without its very central part; see
Sect.\,\ref{subsec_code}) is of interest to us, we consider only the
initial data for $2\,10^8\,{\rm cm} \le r \le 1.2\,10^9\,$cm, and
interpolate all relevant quantities (\eg density, temperature,
composition) onto our Eulerian, lower resolution computational grid
using polynomial interpolation \citep{Press1992}. Due to the
interpolation errors and subtle differences in the input physics, the
interpolated model is no longer in perfect hydrostatic equilibrium. In
order to perfectly balance also the gravitational and pressure forces
in the interpolated model, we use an iterative procedure in the first
hydrodynamic timestep to minimize the numerical fluxes across zone
boundaries. The whole process results in a small temperature decrease
with respect to the temperature profile of the original model
(Fig.\,\ref{fig2.1}). The differences do not exceed a few percent
depending on the radial resolution of the Eulerian grid. The resulting
changes in the density and pressure profiles are negligible due to the
strong electron degeneracy of the gas. The main cause for the slight
de-stabilization of the mapped initial stellar model is the use of
different equations of state in both codes. The hydrodynamic code
employs the equation of state by \citet{TimmesSwesty2000}, whereas the
``Garstec'' code relies on the OPAL equation of state by
\citet{Rogers1996}. At a given density, temperature, and composition
in the helium core during the flash, these equations of state give
pressure values which differ typically by 1 \% the difference being
most apparent in regions where the matter is highly degenerate.

\begin{figure} 
\includegraphics[width=0.99\hsize]{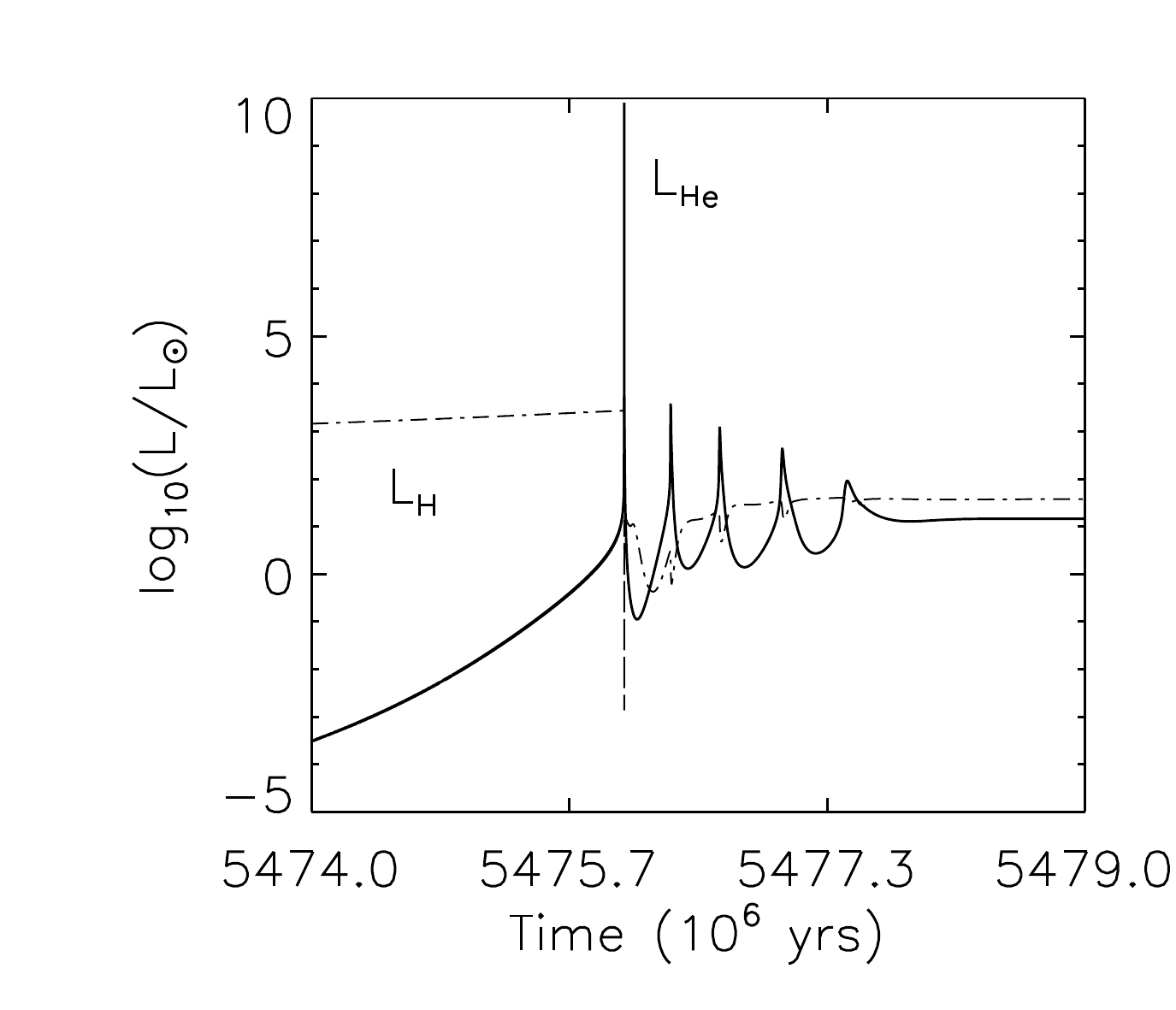} 
\caption{Temporal evolution of the helium luminosity $L_{He}$ (solid)
         versus the hydrogen luminosity $L_{H}$ (dash-dotted) of model
         M during the core helium flash.  }
\label{fig2.4}
\end{figure}

Given that the maximum temperature in the helium core is $T\sim
1\,10^{8} K$, the stellar model reaches the peak in nuclear energy
production rate during the core helium flash in less than $10^{4}$
yrs. The rate at which the nuclear energy production rises is highly
non-linear.  From the onset of the core helium flash at a helium
luminosity $L_{He}\sim 10^{1} \Lsun$, it takes almost 30000\,yrs to
reach $L_{He} \sim 10^{4} \Lsun$, whereas it requires only 40\,yrs to
reach $L_{He} \sim 10^{10}$.  The first core helium flash is followed
by four subsequent mini flashes (Fig.\,\ref{fig2.4}) identified as
thermal pulses by \citet{Thomas1967} until the degeneracy in the
helium core is lifted completely and the star settles down on the
horizontal branch quiescently burning helium in its core.

Since the computed model is a Pop\,I metal rich star, it does not
experience any hydrogen entrainment during the core helium flash
\citep{FujimotoIben1990, SchalttlCassisiSalaris2001}.
 
\section{Input physics and numerics} 
\subsection{Thermal transport} 
The energy flux density due to thermal transport is given by
\begin{equation} 
 f_{\rm cond} = -K_{\rm cond} \nabla T \, , 
\end{equation} 
where $K_{\rm cond}$ is the total conductivity ($\ergcmsK$) and
$\nabla T$ the temperature gradient.

In the helium core, which is partially degenerate, thermal transport
due to both radiative diffusion and electron conduction is important, while
heat transport by ions is negligible, \ie
\begin{equation} 
 K_{\rm cond} = K_\gamma + K_e \, . 
\end{equation} 
The radiative conductivity is given by
\begin{equation} 
 K_\gamma = \frac{4ac}3 \frac{T^3}{\kappa_\gamma\rho} \, , 
\end{equation} 
where $\kappa$, $a$, and $c$ are the Rosseland mean of the opacity,
the radiation constant, and the speed of light, respectively. For the
opacity, we use a fit formula due to \citet{Iben1975} which is based
on the work by \citet{CoxStewart1970a,CoxStewart1970b}. It takes into
account the radiative opacity due to Thomson scattering, free-free
(Krames opacity), bound-bound, and bound-free transitions.

For the thermal transport by electron conduction we consider
contributions due to electron-ion, and electron-electron collisions
which are treated according to \citet{YakovlevUrpin1980}, and
\citet{PotekhinChabrier1997}.

\subsection{Neutrino emission} 
The evolutionary time covered by our hydrodynamic simulations is too
short for neutrino cooling to be of importance. The neutrino losses
computed from the analytic fits of \citet{Itoh1996} give a cooling
rate $\dot\epsilon < 10^2\erggs$, or a corresponding decrease of the
maximum temperature by $|\Delta T|<10^{-1}\K$ over the longest
simulations we performed. Hence, cooling by neutrinos was neglected.

\subsection{Equation of state} 
The equation of state employed in our hydrodynamic code includes
contributions due to radiation, ions, electrons, and positrons. Thus,
the total pressure is given by
\begin{equation} 
 P = P_{\gamma} + P_{ion} + P_e + P_p \, , 
\end{equation} 
where 
\begin{equation} 
 P_{\gamma} = \frac{a}{3}T^4 
\end{equation}
is the radiation pressure of a black body of temperature $T$ ($a$ is
the universal radiation constant), and
\begin{equation} 
 P_{ion} = \sum_{i} \Re \frac{\rho X_i}{A_i} T = \Re \rho T \sum_{i} Y_i\\ 
\end{equation} 
is the pressure of a non-relativistic Boltzmann gas of density $\rho$
consisting of a set of ions of abundance $Y_i=X_i/A_i$ ($X_i$ and
$A_i$ are the mass fraction and the atomic mass number of species $i$,
respectively). $P_e + P_p$ is the pressure of an arbitrarily
degenerate and relativistic electron-positron gas based on table
interpolation of the Helmholtz free energy \citep{TimmesSwesty2000}.

\subsection{Nuclear burning} 
The energy generation rate by nuclear burning is given by
\begin{equation} 
 \dot \varepsilon_{\rm nuc} = \sum_i \frac{\Delta m_i c^2}{m_u} \dot Y_i   
\end{equation}
where
\begin{equation} 
 \Delta m_i = M_i - A_i m_u \, . 
\end{equation} 
is the mass excess of a nucleus of mass $M_i$, and $m_u$ is the atomic
mass unit.
 
Abundance changes are described by a nuclear reaction network
consisting of the four $\alpha$-nuclei $^4$He, $^{12}$C, $^{16}$O, and
$^{20}$Ne, coupled by seven reactions (including the triple-$\alpha$
reaction). We used the reaction rate library of Thielemann (private
communication), which gives the product of the Avogadro number $N_A$
and the velocity averaged cross section $\sv$ in terms of the fit
formula
\begin{eqnarray} 
 N_A\sv = \sum_{l=1}^{n_l}\exp
           \Bigl[c_{1l} &+& c_{2l}T^{-1}+c_{3l}T^{-1/3}+c_{4l}T^{1/3} 
\nonumber\\
                        &+& c_{5l}T+c_{6l}T^{5/3}+c_{7l}\ln T \Bigr]~~, 
\end{eqnarray} 
with rate dependent coefficients $c_{il}$ ($1\le i \le7$). Up to three
sets of coefficients (\ie $1 \le n_l \le 3$) are used. The total
reaction rate due to all one body, two body, and three body
interactions has the form \citep{Mueller1998}:
\begin{eqnarray}
\dot{Y}_{i} = \sum_{j}   c_{i}      \lambda_{j}        Y_{j} 
           &+&\sum_{j,k} c_{i}(j,k) \rho N_A \sv_{j,k} Y_{j} Y_{k} \nonumber\\
           &+&\sum_{j,k,l} 
                   c_{i}(j,k,l)\rho^{2}N^{2}_{A}\sv_{j,k,l} Y_{j} Y_{k} Y_{l}~~,
\label{eq.reaction}
\end{eqnarray}
where the weight factors $c_{i}$ inhibit multiple counts in the sums
over the nuclei {\it{j,k,l}}. The following nuclear reactions were
considered:
\begin{flushleft}
\centerline{
\begin{tabular}{cccccccccc} 
$He^{ 4}$ &+& $C^{12}$ &$\rightarrow$& $ O^{16}$ &+& $\gamma$  & &\\
$He^{ 4}$ &+& $O^{16}$ &$\rightarrow$& $Ne^{20}$ &+& $\gamma$  & &\\
$ O^{16}$ &+& $\gamma$ &$\rightarrow$& $He^{ 4}$ &+& $ C^{12}$ & &\\ 
$Ne^{20}$ &+& $\gamma$ &$\rightarrow$& $He^{ 4}$ &+& $ O^{16}$ & &\\
$ C^{12}$ &+& $C^{12}$ &$\rightarrow$& $Ne^{20}$ &+& $He^{ 4}$ & &\\
$He^{ 4}$ &+& $He^{4}$ &+& $He^{4}$  &$\rightarrow$& $ C^{12}$ &+&
                                                               $\gamma$ \\
$ C^{12}$ &+& $\gamma$ &$\rightarrow$& $He^{4}$  &+& $He^{ 4}$ &+& $He^{4}$ 
\end{tabular}}
\end{flushleft}
Mathematically this results in a nuclear reaction network consisting
of seven non-linear first order differential equations of the form
given by Eq.\,\eqref{eq.reaction} and a temperature equation 
\begin{equation}
  \frac{\partial T}{\partial t} = \dot{\varepsilon}_{nuc} 
                                  \frac{\partial T}{\partial \varepsilon} \, ,
\end{equation}
where $\varepsilon$ is the specific internal energy.

The effects of electron screening were included according to
\citet{DewittGraboske1973} for the triple-$\alpha$ reaction rate, and
in the weak screening regime only.

\subsection{Evolutionary equations} 
The hydrodynamic and thermonuclear evolution of the core helium flash
was computed by solving the governing set of fluid dynamic equations
in spherical coordinates on an Eulerian grid. Using vector notation 
these equations have the form,

\begin{align}
\frac{\partial\,{\bf{U}}}{\partial t} + \nabla\,{\bf{F}} = {\bf{S}}
\end{align}

\noindent
with the state vector {\bf{U}}%

\begin{equation}
{\bf{U}} \equiv \left( \begin{array}{c}
                    \rho \\
                    \rho {\bf{v}} \\
                    \rho e  \\     
                    \rho Y_{i}                 
                 \end{array}\right)
\end{equation}

\noindent
the flux vector {\bf{F}} 

\begin{equation}
{\bf{F}} \equiv \left( \begin{array}{c}
                    \rho {\bf{v}} \\
                    \rho {\bf{v}}{\bf{v}}  \\
                    (\rho e + p){\bf{v}} + f_{cond} \\  
                    \rho Y_{i} {\bf{v}}                     
                 \end{array}\right)
\end{equation}

\noindent
and the source vector {\bf{S}}

\begin{equation}
{\bf{S}} \equiv \left( \begin{array}{c}
                  0 \\
                  - \rho \nabla \Phi  \\
                  - \rho {\bf{v}}\cdot \nabla \Phi + \rho \dot{\epsilon}_{nuc}\\
                  \rho \dot{Y_{i}}                        
                 \end{array}\right)
\end{equation}

\noindent
with $i=1,\ldots,N_{\rm nuc}$ where $N_{\rm nuc}$ is the number of 
nuclear species considered in the nuclear reaction network, and $\rho$,
p, {\bf{v}} and $\Phi$ are the density, pressure, velocity and gravitational 
potential. respectively. The term $f_{cond}$ describes energy transport 
by thermal conduction (see Sect.\,3.1), and $\dot{\epsilon}_{nuc}$ and 
the $\dot{Y_{i}}$ are the nuclear energy generation rate and the
change of the mass fraction of species $i$ due to nuclear reactions, 
respectively (see Sect.\,3.4). The total energy density 
$\rho e = \rho \varepsilon + \rho${\bf{vv}}/2 with $e$ being the 
specific total energy.

\subsection{Code} 
\label{subsec_code}
The numerical simulations were performed with a modified version of
the hydrodynamic code Herakles \citep{Kifonidis2003, Kifonidis2006},
which is a descendant of the code Prometheus developed by Bruce
Fryxell and Ewald M\"uller \citep{MuellerFryxellArnett1991,
FryxellArnettMueller1991}. The hydrodynamic equations are integrated
to second order accuracy in space and time using the dimensional
splitting approach of \citet{Strang1968}, the PPM reconstruction
scheme \citep{ColellaWoodward1984}, and a Riemann solver for real
gases according to \citet{ColellaGlaz1984}. The evolution of the
chemical species is described by a set of additional continuity
equations \citep{PlewaMueller1999}. Source terms in the evolutionary
equations due to self-gravity and nuclear burning are treated by means
of operator splitting. Every source term is computed separately, and
its effect is accounted for at the end of the integration step. The
viscosity tensor is not taken into account explicitly, since the
solution of the Euler equations with the PPM scheme corresponds to the
use of a sub-grid scale model that reproduces the solution of the
Navier-Stokes equations reasonably well \citep{MeakinArnett2007}.
Thermal transport is treated in a time-explicit fashion when
integrating the evolutionary equations.  Self-gravity is implemented
according to \citet{MuellerSteimnetz1995}, while the gravitational
potential is approximated by a one-dimensional Newtonian potential
which is obtained from the spherically averaged mass distribution.
The nuclear network is solved with the semi-implicit Bader-Deufelhard
method which utilizes the Richardson extrapolation approach and
sub-stepping techniques \citep{BaderDeuflhard1983, Press1992} allowing
for very large effective time steps.

The code is vectorized and allows for an adjustment of the vector
length to the memory architecture. Therefore, an optimal performance
on both vector and super-scalar, cache-based machines can be achieved.
   
A program cycle consists of two hydrodynamic timesteps and proceeds
as follows:
\begin{enumerate}
\item The hydrodynamic equations are integrated in $r$-direction
(r-sweep) including the effects of heat conduction. The time averaged
gravitational forces are computed, and the momentum and the total
energy are updated to account for the gravitational source terms.
Subsequently, the equation of state is called to update the
thermodynamic state due to the change of the total energy.
\item Step (1) are repeated in $\theta$-direction ($\theta$-sweep).
\item The nuclear network is solved in all zones with significant
nuclear burning (T $> 10^{8}$K). Subsequently, the equation of state
is called to update the pressure and the temperature.
\item In the subsequent timestep the order of Step (1) and (2) is
reversed to guarantee second-order accuracy of the time integration,
and Step (3) is repeated with the updated quantities.
\item The size of the timestep for the next cycle is determined.
\end{enumerate} 
 
When using spherical coordinates, the CFL stability condition on the
timestep is most restrictive near the origin of the grid. However,
inside a region beneath the off-center temperature maximum there are
no significant non-radial motions to be expected during the evolution
of the core helium flash except in the immediate vicinity of the
temperature inversion, where convective overshooting may occur. Hence,
cutting out the very center of the computational grid does not lead to
any numerical bias, but saves considerable amounts of computational
time. In the radial direction we used a closed (\ie reflective) outer
and inner quasi-hydrostatic boundary obtained by means of polynomial 
extrapolation, which significantly suppresses any artificial velocity 
fluctuations resulting from an imbalance of gravitational and 
pressure forces in the boundary zones. For two-dimensional runs, 
the boundary conditions in the angular direction are reflective as well.

\begin{figure}
\includegraphics[width=0.99\hsize]{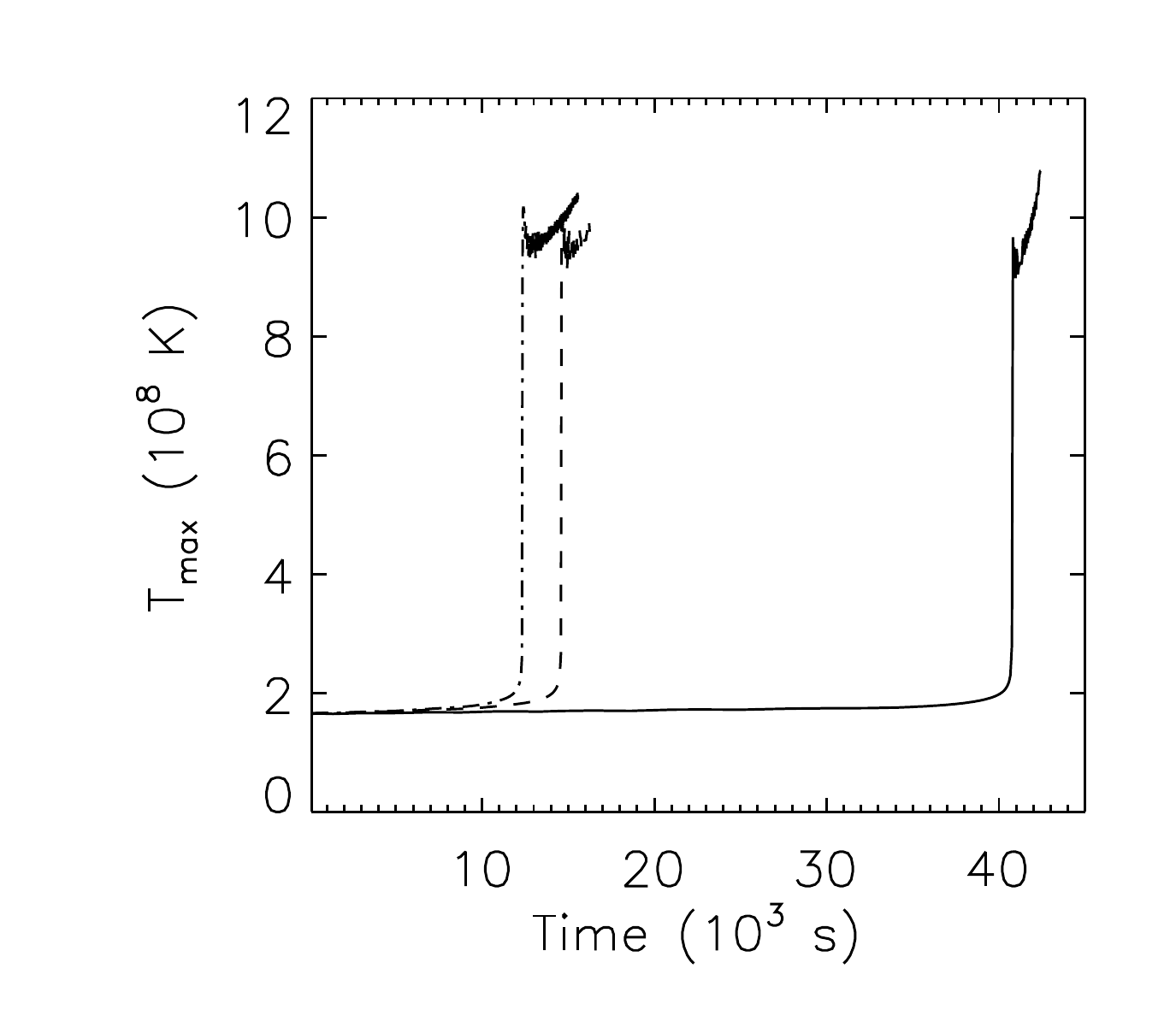} 
\caption{Evolution of the temperature maximum $T_{max}$ in the
         one-dimensional models JE2 (solid), JE3 (dashed),
         and JE4 (dash-dotted), respectively.}
\label{fig4.1.2a}  
\end{figure} 

\begin{figure}
\includegraphics[width=0.99\hsize]{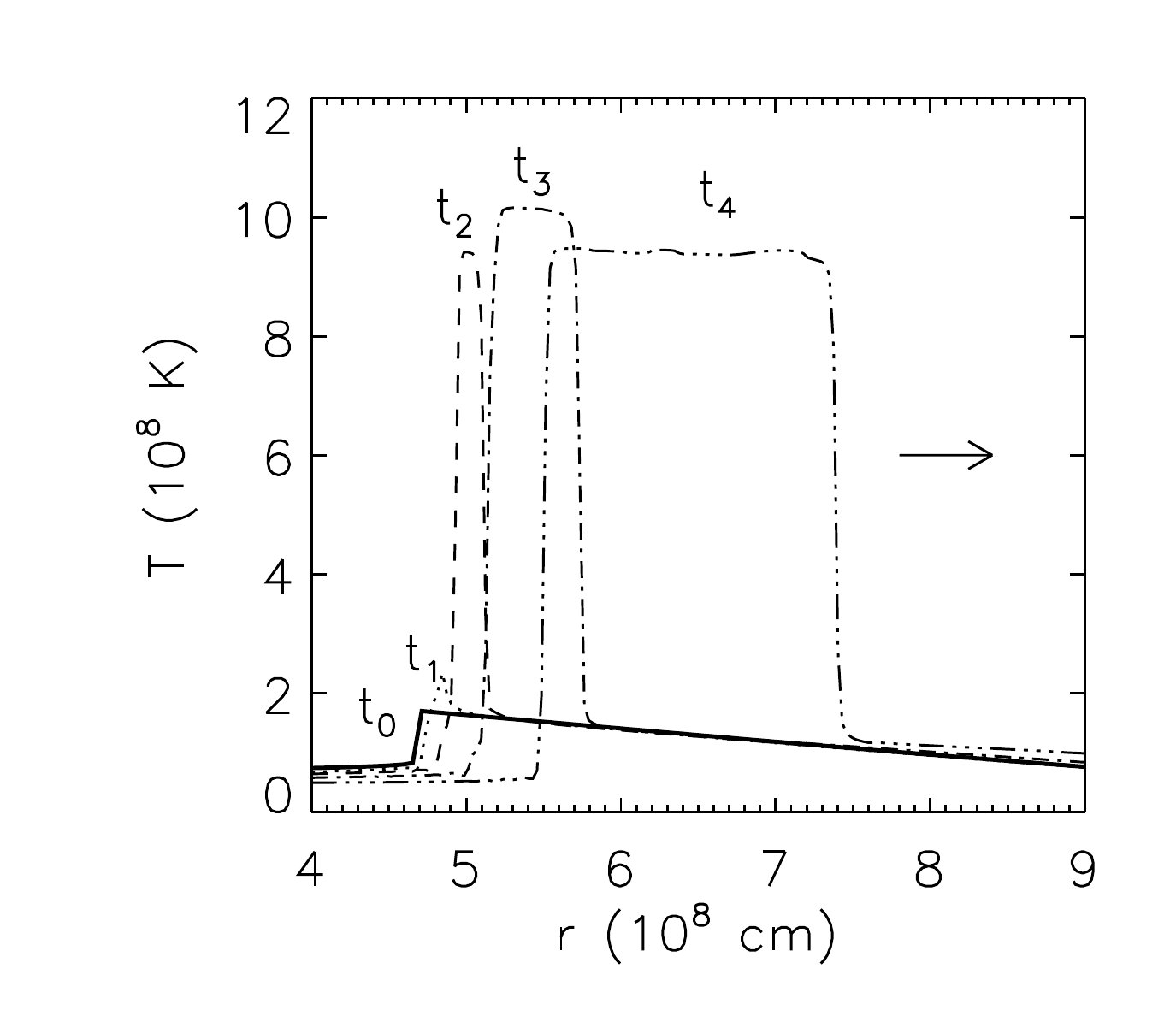}
\caption{Temperature stratification across the helium core in model
         JE4 during the runaway at $t_{1}$ = $12270$~s (dotted),
         $t_{2}$ = $12352$~s (dashed), $t_{3}$ = $12392$~s
         (dash-dotted), and $t_{4}$ = $12762$~s (dash-dot-dotted),
         respectively.  The solid line corresponds to the initial
         model ($t_{0}$), and the arrow indicates the direction of the
         flame propagation.}
\label{fig4.1.2b}  
\end{figure}

After interpolation and stabilization, the initial model in the
two-dimensional simulations had to be perturbed explicitly to trigger
convection, because an initially exactly spherically symmetric model
remains that way for ever when evolved in spherical coordinates with
our code. We imposed a random flow field with a maximum (absolute)
velocity of $10 \cms$, and random density perturbations with $\Delta
\rho / \rho \le 10^{-2}$.
 
\section{Results of 1D simulations} 

\begin{table}[!ht] 
\caption{Some properties of the 1D simulations: number of radial grid
         points ($N_{r}$), radial resolution ($\Delta r$ in
         $10^{8}$cm), time up to the thermonuclear runaway, $t_{trn}$,
         and maximum evolution time $t_{max}$ (both in s).}
\begin{center}
\begin{tabular}{p{1.cm}|p{1.cm}p{1.cm}p{1.cm}p{1.cm}} 
\hline
\hline
run & $N_{r}$ & $\Delta r$ & $t_{trn}$ & $t_{max}$ 
\\
\hline
JE2  & 180  & 5.55 & 40700 & 42500 \\ 
JE3  & 270  & 3.77 & 14600 & 16250 \\ 
JE4  & 360  & 2.77 & 12300 & 15600 \\ 
\hline
\end{tabular}
\end{center} 
\label{simu1d} 
\end{table}   

We have performed several one-dimensional simulations using model M,
which differ only by their grid resolution (see
Table\,\ref{simu1d}) to see whether without allowing for convective
flow a thermonuclear runaway can be avoided.

Figure\,\ref{fig4.1.2a} demonstrates that heat conduction and
adiabatic expansion alone fail to stabilize the model, \ie
one-dimensional hydrodynamic simulations result in a thermonuclear
runaway.  Initially, the maximum temperature increases only slowly,
but it starts to rise rapidly after a time $t_{trn}$
(Tab.\,\ref{simu1d}) up to a value $T \sim 10^9\K$. For instance, from
the temperature evolution of model JE4 one can determine that a local
hot spot with a temperature of $2.3\,10^8\K$ will runaway after about
$80\,\s$ (Fig.\,\ref{fig4.1.2b}). The time at which the runaway is
triggered depends on the grid resolution, being longer in models with
lower resolution (Fig.\,\ref{fig4.1.2a}).

In every case, a thermonuclear flame with $T \sim 10^{9}\,$K
ultimately forms and propagates outwards with a subsonic velocity
depending on the grid resolution. Since our two-dimensional (more
realistic) simulations do not show such a behavior, we will refrain
from further discussing details of the one-dimensional simulations.

\section{Results of 2D simulations} 
\begin{table*} 
\caption{Some properties of the 2D simulations: number of grid points
         in radial ($N_{r}$) and angular ($N_{\theta}$), radial
         ($\Delta r$ in $10^{8}$cm) and angular grid resolution
         ($\Delta \theta$), characteristic length scale $l_{c}$ of the
         flow (in $10^{8}$cm), characteristic velocity $v_{c}$ of the
         flow (in $10^{6}\cms$), Reynolds number $R_{n}$ associated
         with the numerical viscosity of our code
         \citep{PorterWoodward1994}, damping time-scale due to the
         numerical viscosity $t_{n}$, typical convective turnover time $t_{o}$,
         and maximum evolution time $t_{max}$ (in s), respectively.}
\begin{center}
\begin{tabular}{p{1.cm}|p{1.5cm}p{1.0cm}p{1.0cm}p{1.0cm}
  p{1.0cm}p{1.0cm}p{1.0cm}p{1.cm}p{1.cm}p{1.cm}p{1.cm}p{1.cm}p{1.cm}} 
\hline
\hline
run & grid & $\Delta r$ & $\Delta\theta$ & $l_{c}$ & $v_{c}$ 
& $R_{n}$ & $t_{n}$& $t_{o}$ & $t_{max}$ 
\\
\hline 
DV2 & 180$\times$90 & 5.55 & 2\degr & 
4.7 & 1.03  & 1900 & 11000 & 910 & 30000  \\ 
DV3 & 270$\times$180 & 3.70 & 1\degr & 
4.7 & 1.46  & 8900 & 36000 & 640 & 30000 \\
DV4 & 360$\times$240 & 2.77 & 0.75\degr & 
4.7 & 1.52 & 21000 & 83000 & 620 & 30000 \\ 
\hline
\end{tabular} 
\end{center}
\label{simu2d} 
\end{table*}  

In Table\,\ref{simu2d} we summarize some characteristic parameters of
our two-dimensional simulations that are based on model M.

We will first discuss one specific simulation DV4 in some detail,
which serves as a standard to which we will compare the results of
other runs. Thereafter, we will discuss some general properties of all
2D simulations. Every simulation covered approximately $30000\s$
($\sim 8\,$hrs) of the evolution near the peak of the core helium
flash. They were performed on an equidistant spherical grid
encompassing 95\% of the helium core's mass (X($^{4}$He)$>$0.98)
except for a central region with a radius of $r = 2\,10^8\cm$ which
was excised in order to allow for larger timesteps. As this radius is
sufficiently smaller than the radius of the temperature inversion ($r
\sim 5\, 10^8\cm$), its presence does not influence the convection
zone.

\begin{figure*}
\includegraphics[width=0.49\hsize]{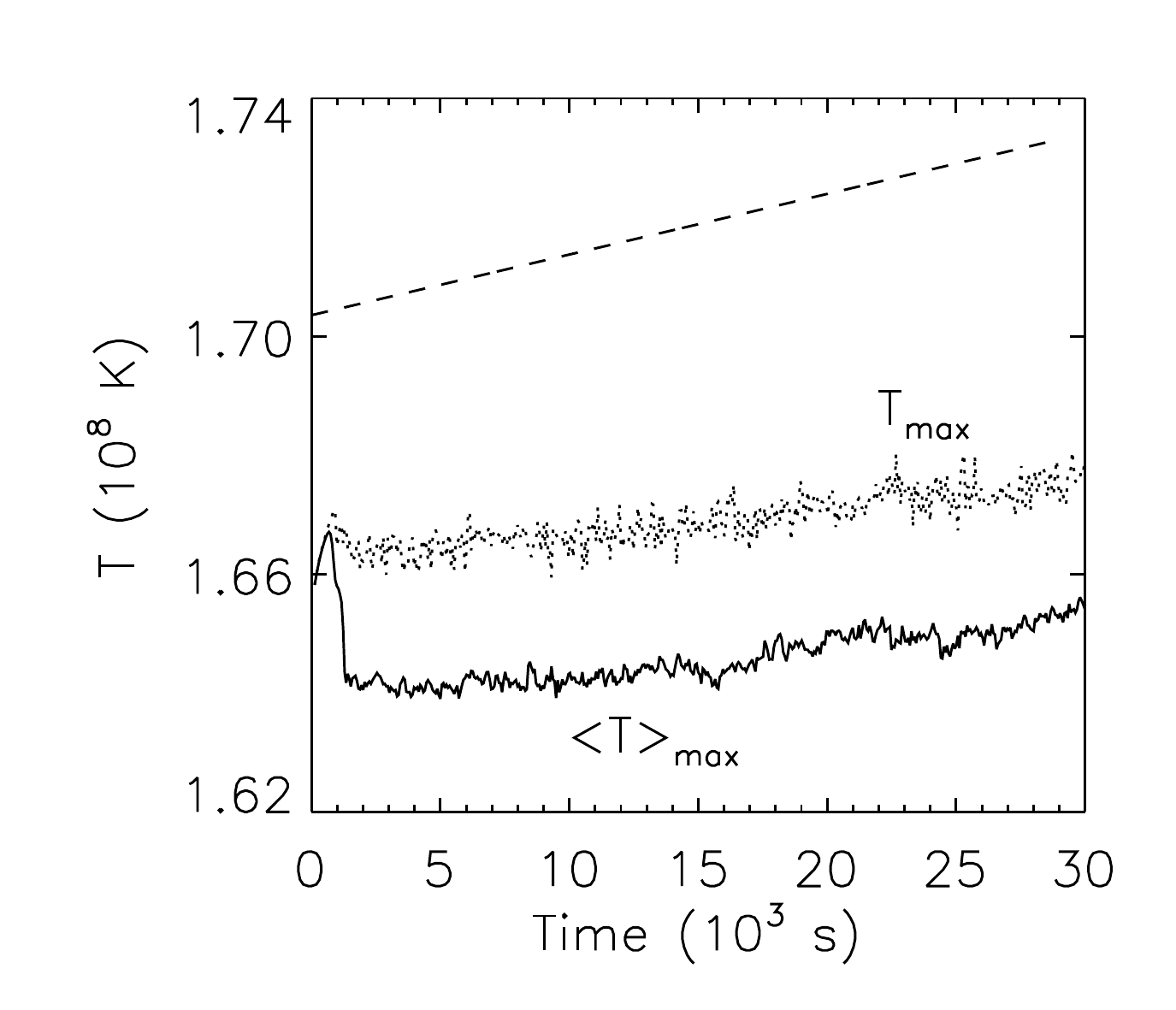}
\includegraphics[width=0.49\hsize]{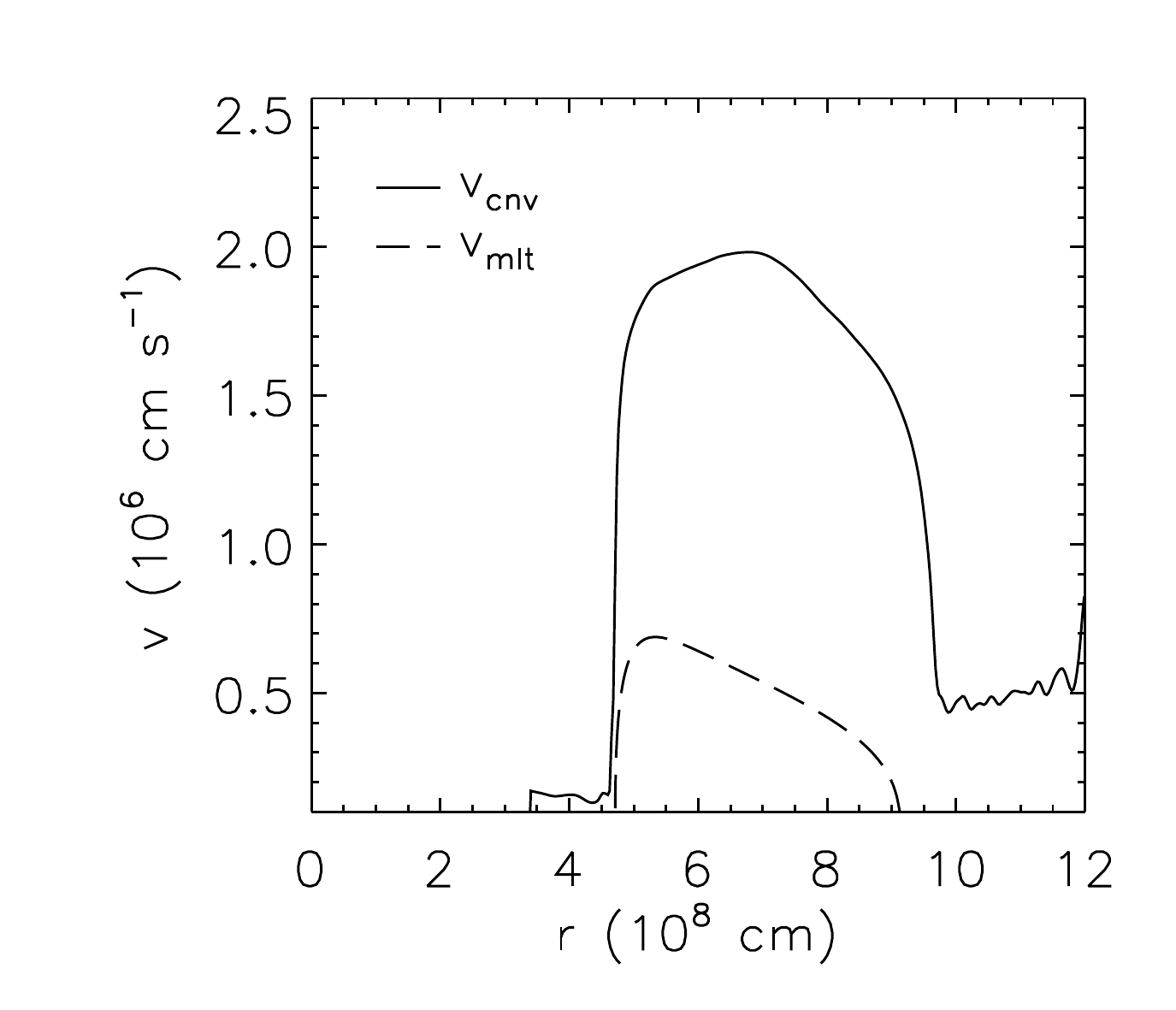}
\caption{{\it{Left panel:}} Temporal evolution of the horizontally
          averaged temperature maximum $\langle T \rangle_{max}$
          (solid), and of the global temperature maximum $T_{max}$
          (dotted) in model DV4. The dashed line corresponds to the temporal
          evolution of the maximum temperature in the stellar
          evolutionary calculations of the model M.  {\it{Right
          panel:}} The r.m.s convection velocity v$_{cnv}$ in
          simulation DV4 averaged over 6000\,s (solid) versus the
          convection velocity predicted by the mixing length theory
          v$_{mlt}$ (dashed).}
\label{fig5.1.2}  
\end{figure*}

\subsection{Simulation DV4} 

\begin{figure*}
\includegraphics[width=6.3cm]{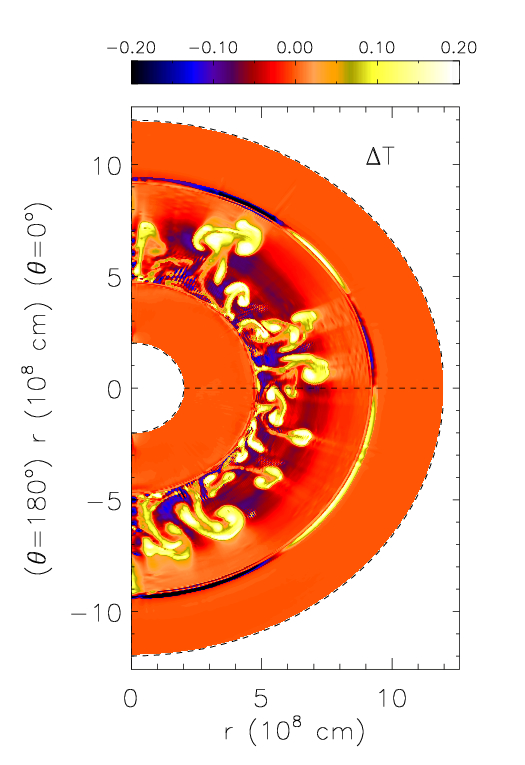}
\includegraphics[width=6.3cm]{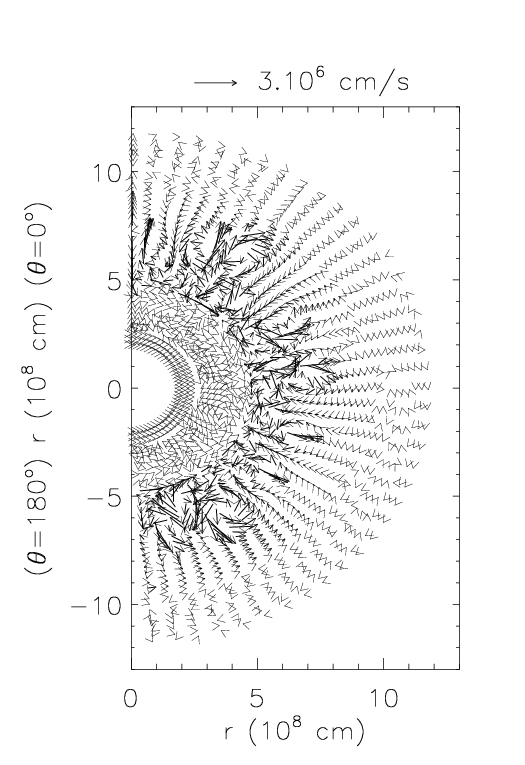} 
\includegraphics[width=6.3cm]{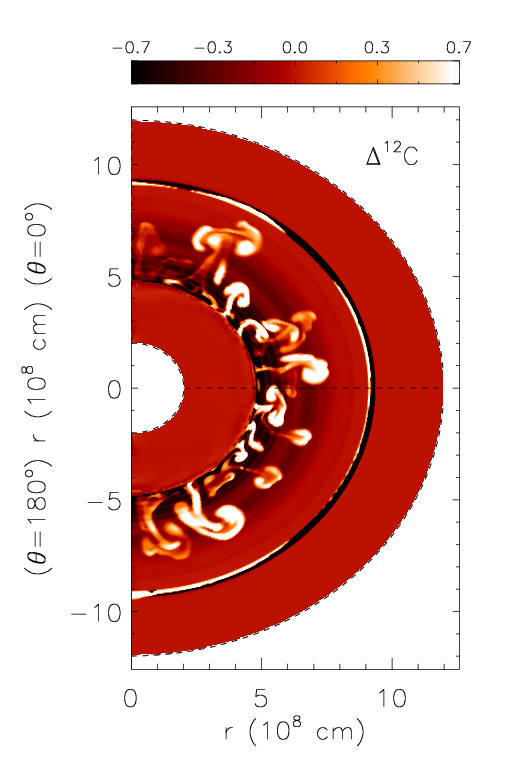} 

\vspace{0.25cm}

\includegraphics[width=6.3cm]{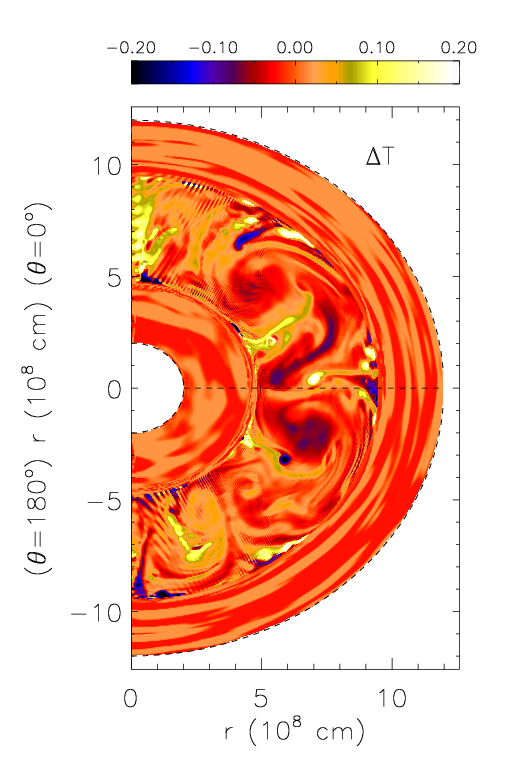}
\includegraphics[width=6.3cm]{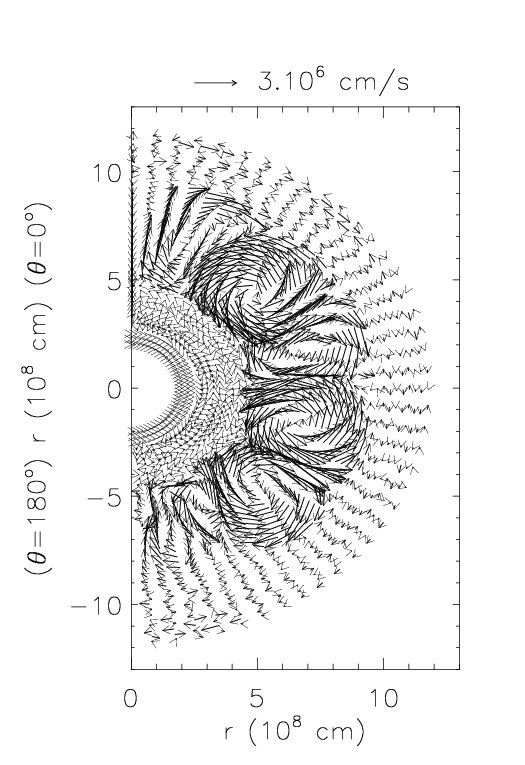} 
\includegraphics[width=6.3cm]{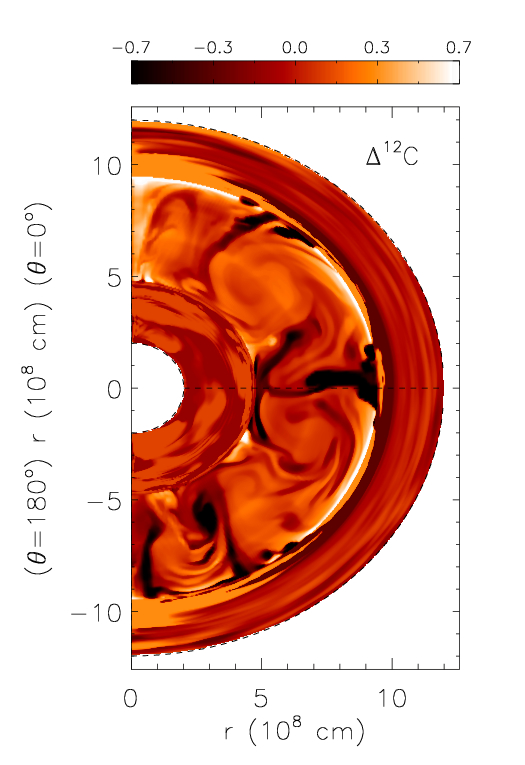}
\caption{Snapshots of the onset of convection at $1020\,$s (upper
         panels), and of the evolved convection at $29000\,$s (lower
         panels) in model DV4, showing the temperature contrast
         $\Delta T = \mbox{100}\, (T - \langle T \rangle_{\theta}) /
         \langle T \rangle_{\theta}$ (left panels), the velocity field
         (middle panels), and the $^{12}C$ contrast $\Delta ^{12}C =
         \mbox{100}\, (^{12}C - \langle^{12}C \rangle_{\theta}) /
         \langle^{12}C\rangle_{\theta}$ (right panels), respectively.
         $\langle \rangle_{\theta}$ denotes a horizontal average at a
         given radius.}
\label{fig5.3}
\end{figure*}

After the start of the simulation the initial velocity perturbations
begin to grow in a narrow layer just outside the temperature maximum
($r\sim5\,10^{8}\cm$),
\ie in the region heated by nuclear burning.  Later on at t $\sim
800\,s$, several hot bubbles appear, which rise upward with maximum
velocities $\sim 4\,10^{6}\,\cms$ (Fig.\,\ref{fig5.3}). They are
typically about 0.2\% hotter than the angular averaged temperature at
a given radius. The $^{4}$He mass fraction of all hot bubbles is about
0.4\% less than the corresponding angular averaged value since helium
has been depleted in the bubbles by the triple\,$\alpha$ reaction.
Consequently, $^{12}$C and $^{16}$O (produced in helium burning) are
enhanced by $\sim$ 0.7\% in the bubbles.

During the first $700\,s$ of the evolution, the off-center 
maximum mean temperature $\langle T \rangle_{max}$ rises with a rate 
of $\sim 1000\Ks$ until it reaches a value $\sim 1.67\,
10^{8}\,K$. At this moment, from the region around the 
$\langle T \rangle_{max}$, 
the bubbles emerge and cause its decrease by $\sim 2.6\,
10^{6}\,K$ in just $570\,$s corresponding to a temperature drop rate
of $4540\Ks$ (Fig.\,\ref{fig5.1.2}). This phase marks the onset of
convection where a fraction of the thermonuclear energy released via
helium burning starts to be efficiently transported away from the
burning regions by matter flow, thereby inhibiting a thermonuclear
runaway.

Once the bubbles form, they rise upwards and start to interact and
merge, \ie the convective layer begins to grow in radius.  About $\sim
1300\,$s after the start of the simulation, the whole convection zone
is covered by an almost stationary flow pattern with an almost
constant total kinetic energy of the order of $10^{45}\,$erg.  At this
time vortices dominate the flow pattern. They extend across the whole
convective region ($\sim 2.1 H_{p}$), and are of approximately similar
angular size, one vortex covering about 40 degrees (diameter $\sim 5\,
10^{8}\,$cm). Usually we find about four such vortices with two
dominant up-flows of hot gas at $\theta \sim 60^{\circ}$, and $\theta
\sim 120^{\circ}$, respectively (see, \eg Fig.\,\ref{fig5.3}).  These
large vortices are rather stable surviving until the end of our
simulations. Typical convective flow velocities are $v_{cnv} \sim
1.5\, 10^{6}\,\cms$, and thus well below the local sound speed
($c_{S}\sim 1.7\, 10^{8}\,\cms$), \ie a vortex requires about 600\,s
for one rotation. The persistence of vortices is not typical for
turbulent convection.

The dominance of large scale structures might be a consequence of the
usage of a Riemann solver based compressible code. The Mach number $M$
of the convective flow is $\sim0.01$. Is PPM suited for this kind of
subsonic flow? This question, which is beyond the scope of the present
study, needs to be investigated, as it is know that the artificial
viscosity of standard Riemann solver methods exhibit incorrect 
scaling with the flow Mach number as $M \rightarrow$ 0. 
\citep{Turkel1999} \ie the
inherent artificial viscosity of PPM may be too high for adequately
simulating flows at low Mach numbers (\eg $M \sim 0.01$).

Energy transport by convection within the vortices is concentrated
into a few narrow upward drafts, compensated partially, but only to a
small extent, by down-flows.  The vortices transport energy mostly
along their outer edges.  Matter in their centers does not interact
with regions of dominant nuclear energy production at all.

The horizontally averaged value of the maximum temperature, barring
some additional temperature fluctuations due to convection, is
slightly rising after the onset of convection during the whole
subsequent evolution with a rate of around $40\,\Ks$ (see
Fig.\,\ref{fig5.1.2}). This rate seems to be about 60\% smaller than
the rate seen in the stellar evolutionary calculations ($\sim
100\,\Ks$), which could be either a result of the initially lower value of
the temperature maximum after the stabilization phase at the beginning
of the simulation (see Sec.\,2) or more dynamic convective
motion, since the mean convective velocities $v_{cnv}$ exceed the velocities
predicted by mixing length theory, v$_{mlt}$, on average by a factor 
of four (Fig.\,\ref{fig5.1.2}).

Convection distributes the energy in such a way that the temperature
gradient $\nabla$ never significantly exceeds $\nabla_{ad}$ in model
M. Although, the value of $\nabla$ established at the beginning of the
simulation deviates slightly after some time from the gradient at
later times, it remains close to the adiabatic temperature gradient
$\nabla_{ad}$ (the relative difference is less than 1\%). In this
respect there is thus no indication of any significant deviation from
the situation obtained in stellar evolutionary calculations. 

\begin{figure*} 
\includegraphics[width=0.49\hsize]{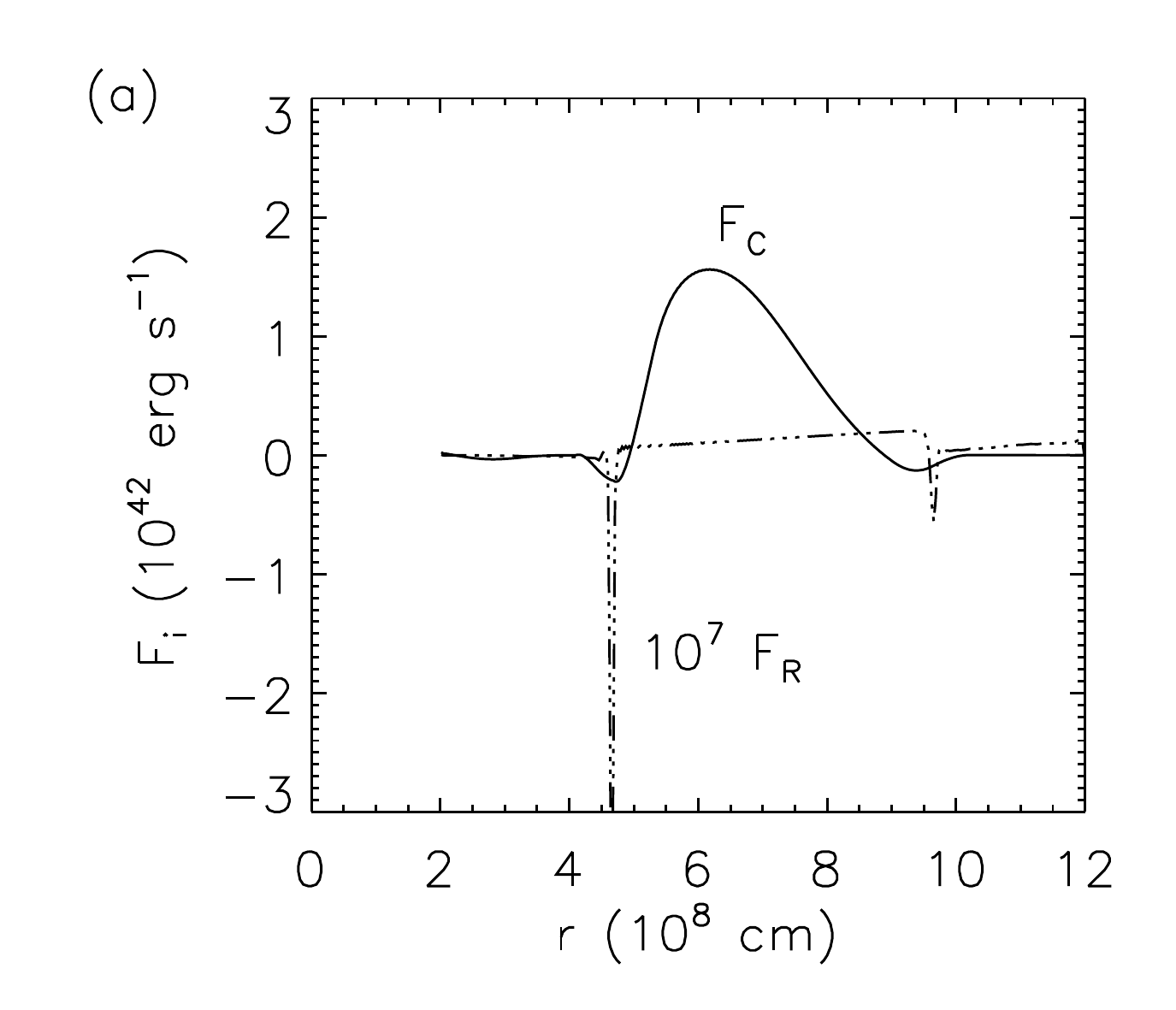} \hfill
\includegraphics[width=0.49\hsize]{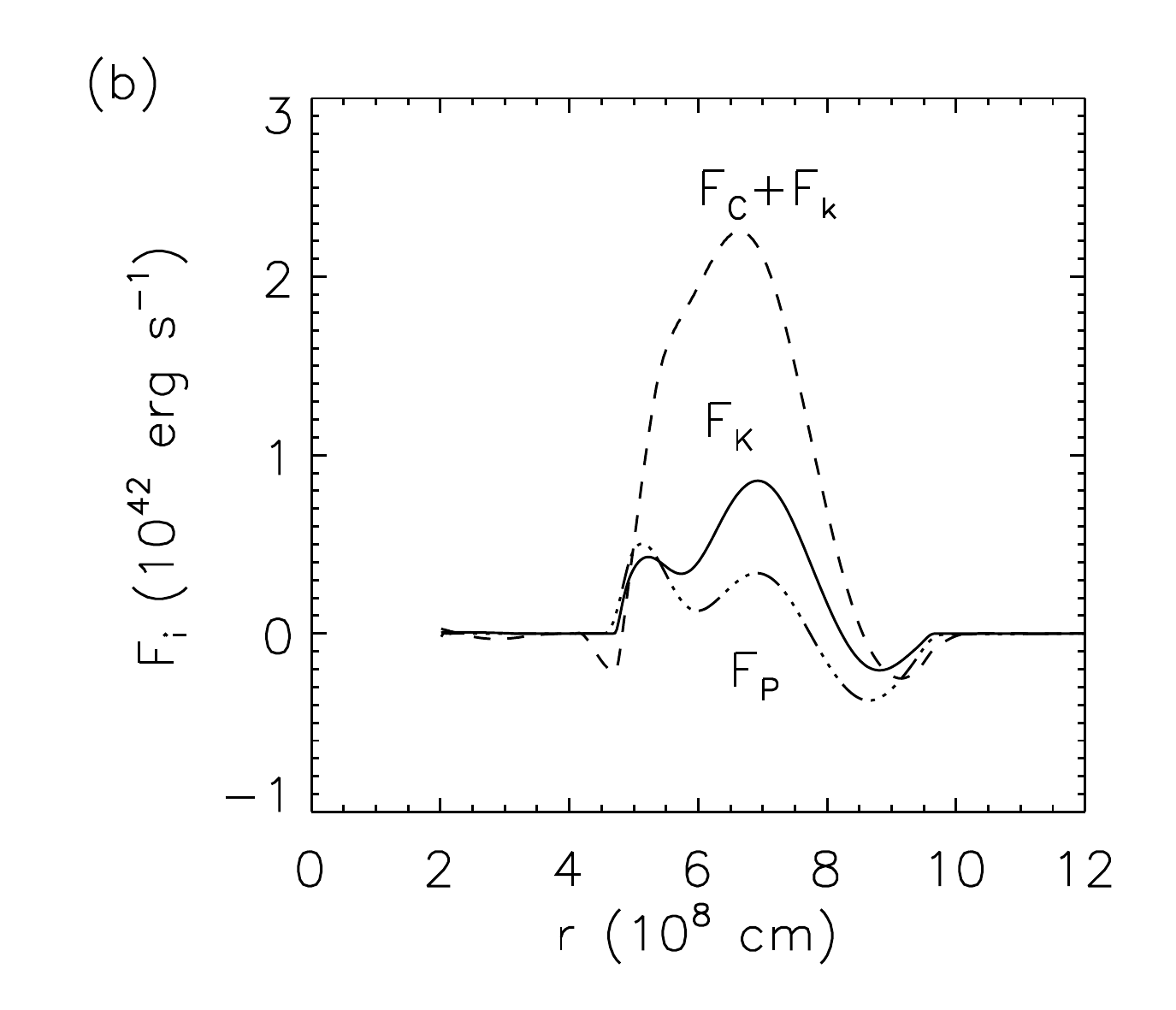} \vskip.2cm  
\includegraphics[width=0.49\hsize]{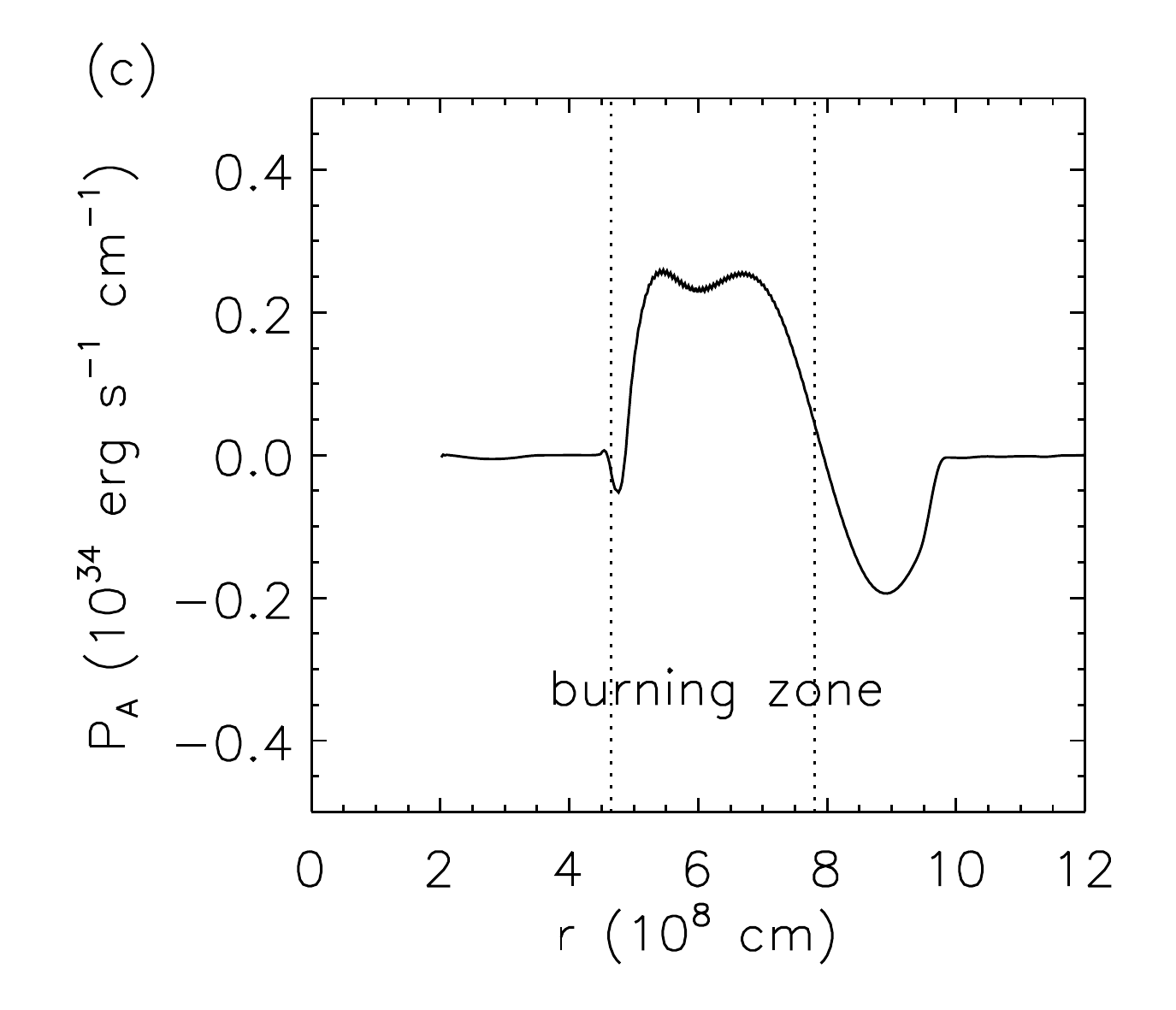} \hfill
\includegraphics[width=0.49\hsize]{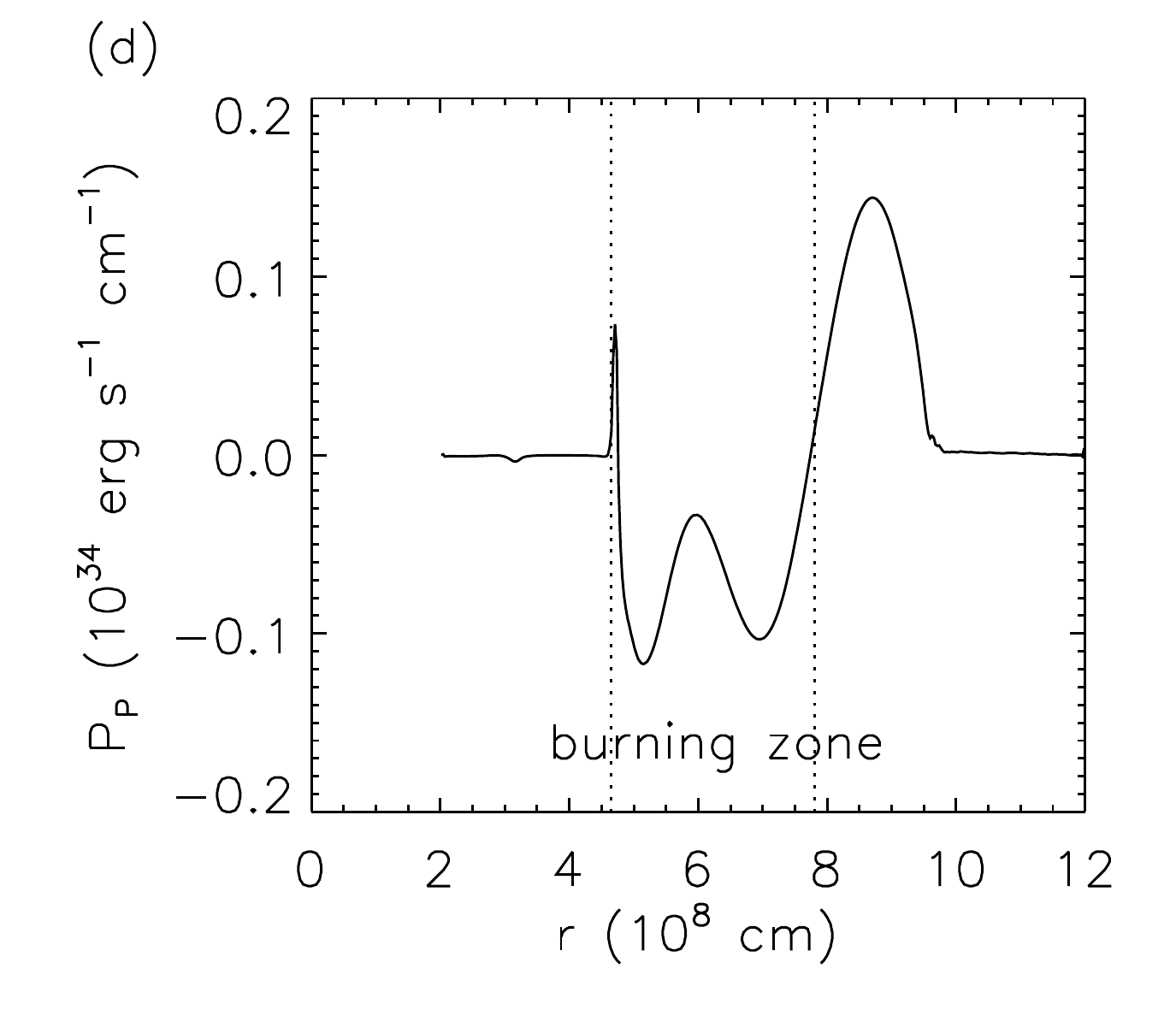}
\caption{Snapshots of various energy fluxes and source terms in model
         DV4 (time averaged over $6000\,$s from $t = 18000\,$s to $t =
         24000\,$s): (a) convective flux $F_C$ (solid), and the energy
         flux due to the thermal transport $F_{R}$ (dash-dotted); (b)
         kinetic flux $F_K$ (solid), acoustic flux $F_P$
         (dash-dot-dotted), and sum of the kinetic and convective flux
         $F_{C}+F_{K}$ (dashed); (c) source terms due to work done by
         buoyancy forces $P_{A}$, and (d) due to volume changes
         $P_P$. The vertical lines enclose the nuclear burning zone
         (T$\,> 10^{8}$ K).}
\label{fig5.5}
\end{figure*} 

The apparent spike in the initial $^{12}C$ distribution at the
location of the temperature maximum (Fig.\,\ref{fig2.2.3}) is a result
of a non-instantaneous treatment of the convective mixing in stellar
evolutionary calculations. It turns out that a non-instantaneous
treatment of mixing is not required during the core helium flash since
simulation DV4 indicates that the spike gets smeared out immediately
after convection is triggered. This implies that the assumption of
instantaneous mixing is a good approximation locally, despite the
strong temperature dependence of the energy production rate.

\subsubsection{Energy fluxes}

Fig.\,\ref{fig5.5} displays the individual contributions of various
energy fluxes, time-averaged over many convective turnover times, \ie
only the average effect of convection should be apparent. The
derivation of these quantities is explained in Appendix\,A.  All
energy fluxes, $F$, describe the amount of energy which is transported
per unit of time across a sphere of given radius.

Most of the nuclear energy production in the convection zone takes
place in a relatively narrow shell around the location of the
temperature maximum.  This energy is transported away by both
convection and thermal transport due to heat conduction and radiation.
The convective (or enthalpy) flux, $F_{C}$, varies from $-0.2\,
10^{42}\,\ergs$ up to $1.6\, 10^{42}\,\ergs$.  The kinetic flux,
$F_{K}$, reaches a value of at most $1\, 10^{42}\,\ergs$, and is
mostly positive in the convection zone, \ie the motion has a
predominantly upward direction. This implies that the fast narrow
upward directed streams are dominating over the slower and broader
downward flows. The ratio of the extreme values of $F_{C}$ and $F_{K}$
is nearly 2:1, \ie nuclear energy is mainly stored in the internal
energy of rising hot gas.  

\begin{figure*} 
\includegraphics[width=0.49\hsize]{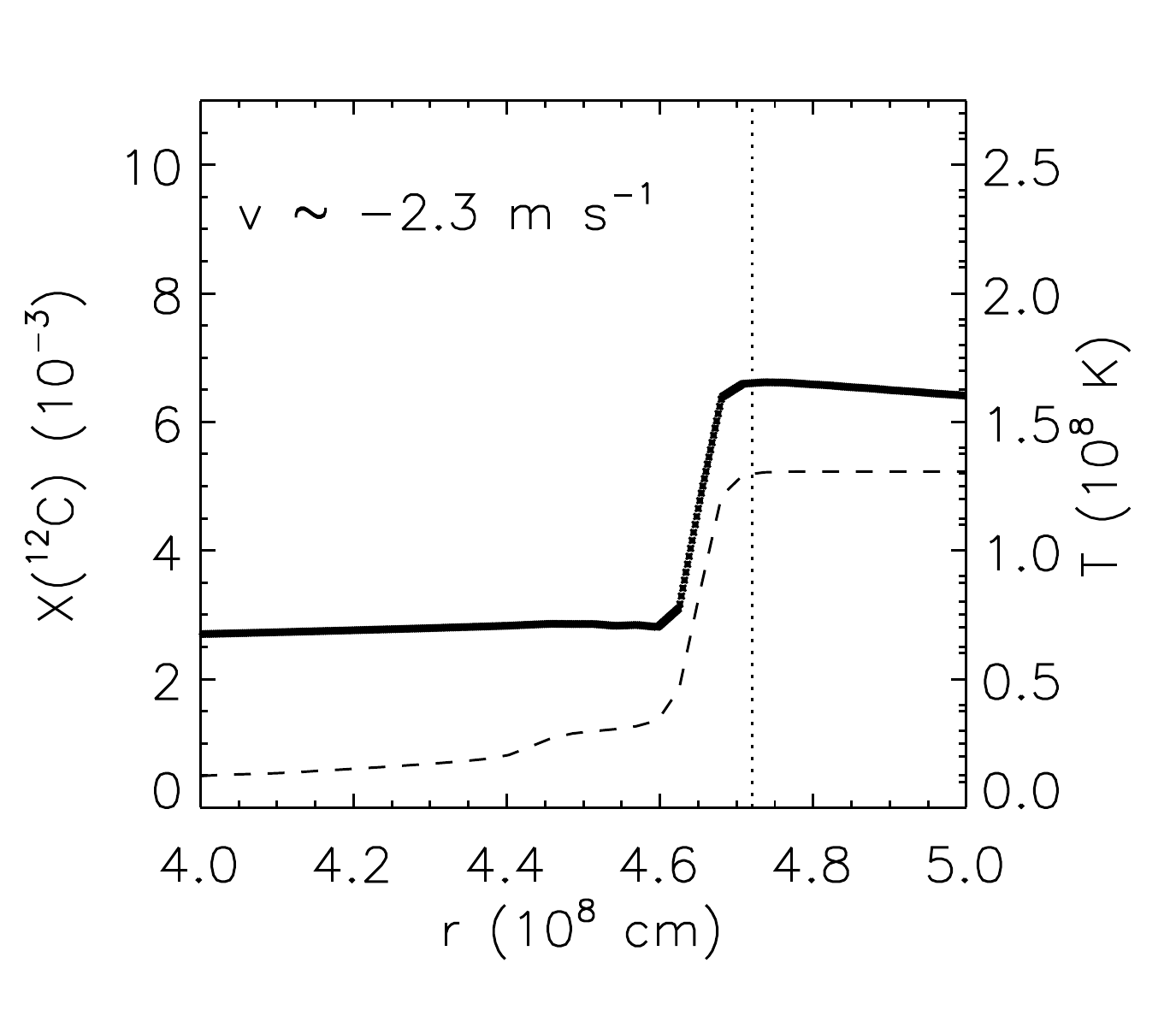}
\includegraphics[width=0.49\hsize]{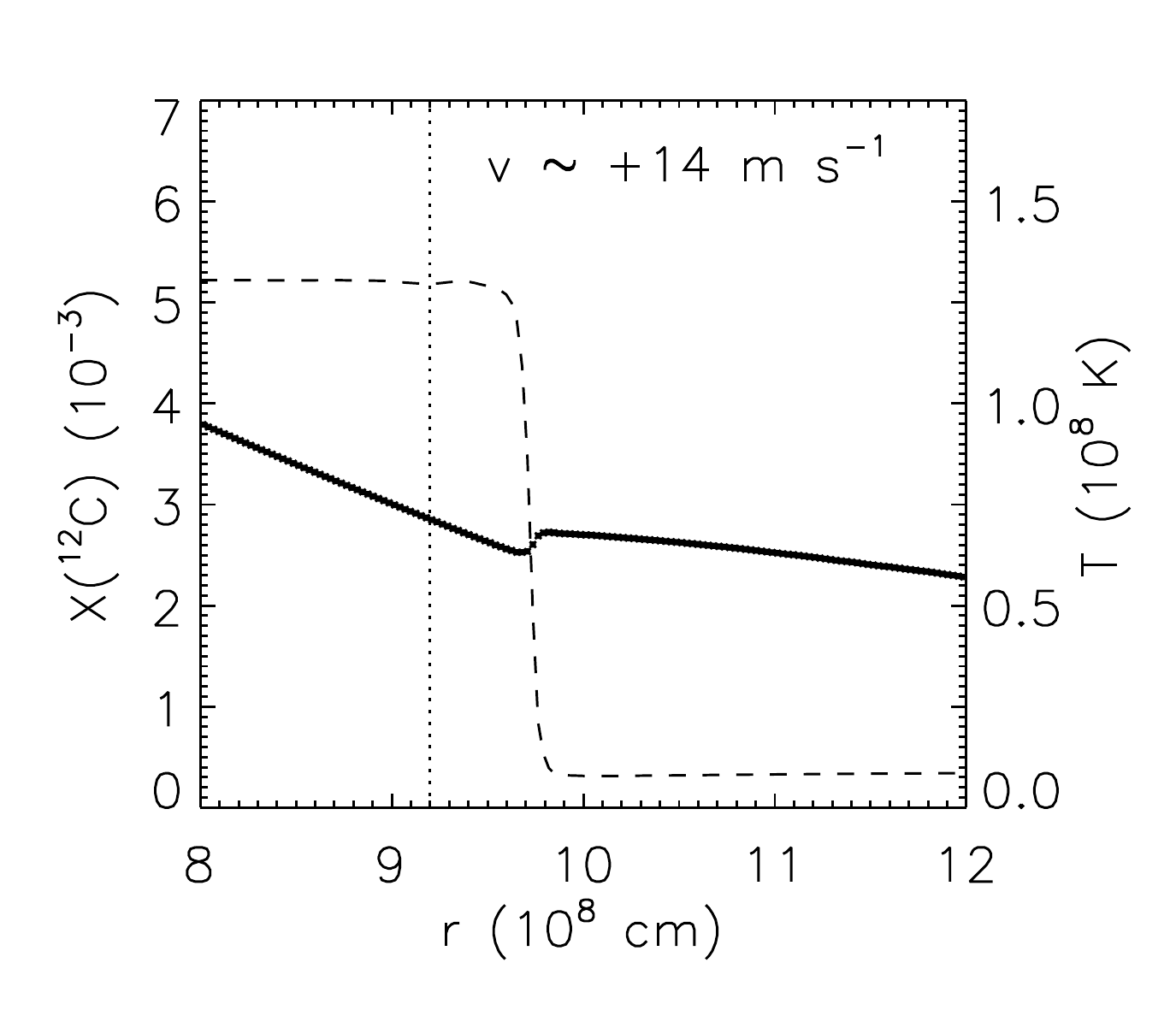}
\caption{Angular averaged $^{12}C$ distribution (dashed) and
         temperature stratification (thick) at the inner (left panel)
         and outer edge (right panel) of the convection zone in model
         DV4 at $t = 30000\s$.  The vertical dotted lines mark the
         initial boundaries of the convection zone at $t = 0\,$s.}
\label{fig5.6.7}  
\end{figure*} 

Convective and kinetic energy flux together
transport more than 90\% of the generated nuclear energy upward
through the convection zone, the value is dropping to zero towards its
border. Part of the heat released in the nuclear processes is in fact
transported downwards towards the inner edge of the temperature
inversion.  Almost none of the nuclear energy reaches the surface of
the helium core, neither by convection nor by conduction, \ie all the
energy released is deposited within the core causing its expansion.
Energy transport due to heat conduction and radiation is everywhere
negligible compared to the other contributions.  The viscous flux,
$F_{V}$, is small as well, and losses due to friction, $P_{V}$, only
influence the dynamics significantly near the borders of the
convection zone \citep{Achatz1995}.

For completeness we also consider the flux and source terms of the
kinetic energy (see Appendix\,A), which allow for a further insight
into the operation of convection.  The radial profile of the source
term $P_{A}$, corresponding to the work done by buoyancy forces, shows
that the vertical convective flows are accelerated due to their
density fluctuations in the entire region of dominant nuclear burning
(burning zone) above $T_{max}$. Corresponding pressure fluctuations
(causing expansion due to a pressure excess, respectively compression
due to a pressure deficit) powered by the volume work $P_{P}$ show
that the gas within the burning region expands, which effectively
again implies that an acceleration occurs. Due to the importance of
$P_{P}$ in the convection zone, the acoustic flux $F_{P}$, which
transports pressure fluctuations, reaches a value comparable to that
of the kinetic flux $F_{K}$, its value being negligible elsewhere.

\subsubsection{Turbulent entrainment, temperature inversion and the 
 growth of the convection zone}
Turbulent entrainment (commonly referred to as overshooting) is a
hydrodynamic process allowing for mixing and heating in regions which
are convectively stable according to the Schwarzschild or Ledoux
criterium. Turbulent entrainment, \ie penetration beyond the formal
convective boundaries, takes place at both edges of the convection
zone, and is driven by down-flows and up-flows.  We study the
entrainment by monitoring the temperature changes and the $^{12}$C
concentration at the (formal) edges of the convection zone.  $^{12}$C
is the most suitable element for investigating the extent of
convective mixing, because at the beginning of the simulations, it is
mostly absent outside the convection zone, and therefore can be
enhanced there only due to overshooting.

At $t = 30000\,$s, \ie near the end of simulation DV4, the temperature
inversion is located at $r=4.65\, 10^{8}$ cm (Fig.\,\ref{fig5.6.7}).
Thus, it is about $70\,$km closer to the center of the star than it
was at the beginning of the simulation ($4.72\, 10^{8}$ cm). Its shape
remains almost unchanged and discontinuous during the whole evolution,
and its propagation speed can be estimated from the heating rate
$\delta T / \delta t \sim 2760\,\Ks$ and the local gradient
$\delta T / \delta r \sim 12\,\Kcm$ at the steepest point of
the inversion:
\begin{equation}
 \mbox{v} \simeq -  (\delta T / \delta t)\,/\,(\delta T / \delta r) \sim -2.3\, \mes 
\end{equation}
This speed is significantly higher than the propagation speed due to
the heat conduction alone. Note that the energy flux carried by the
heat conduction is seven orders of magnitude smaller than the energy
flux carried by the convection.  Assuming that the convective energy
flux at the position of the temperature inversion ($F_{c} \sim 0.2\,
10^{42}\,\ergs$) is used up completely to heat the layers beneath the
temperature inversion, a typical heating rate of $\dot{T} =
\dot{E}/C_{inv} \sim 1250\,\Ks$ can be derived, which is a bit smaller
than the value inferred from the simulation, but still in good
agreement.  $C_{inv}$ is the heat capacity of the layers including the
temperature inversion ($C_{inv} \sim 1.6\, 10^{38}\,\ergK$). This
implies that turbulent entrainment leads to a strong heating of the
inner neutrino cooled center of the star that occurs on timescales
which are relatively short compared to stellar evolutionary
timescales. Such a heating was studied already by
\citet{DeupreeCole1983} and \citet{ColeDemDeupree1985} who obtained
qualitatively similar results.  Note, that in the one-dimensional
stellar evolution calculations the temperature maximum moves outwards
with time.

Assuming that the estimated propagation speed of the temperature
inversion remains constant, it would reach the center of the helium
core and lift the electron degeneracy there in just 24 days. This
scenario would rule out the occurrence of mini-flashes subsequent to
the main core helium flash, which are observed in stellar evolutionary
calculations (Fig.\,\ref{fig2.4}).  Moreover, as in stars with higher
mass and helium abundance the flash occurs closer to the center
\citep{SweigertGross1978}, in these stars the center can be reached
even faster.

\begin{figure*} 
\includegraphics[width=0.49\hsize]{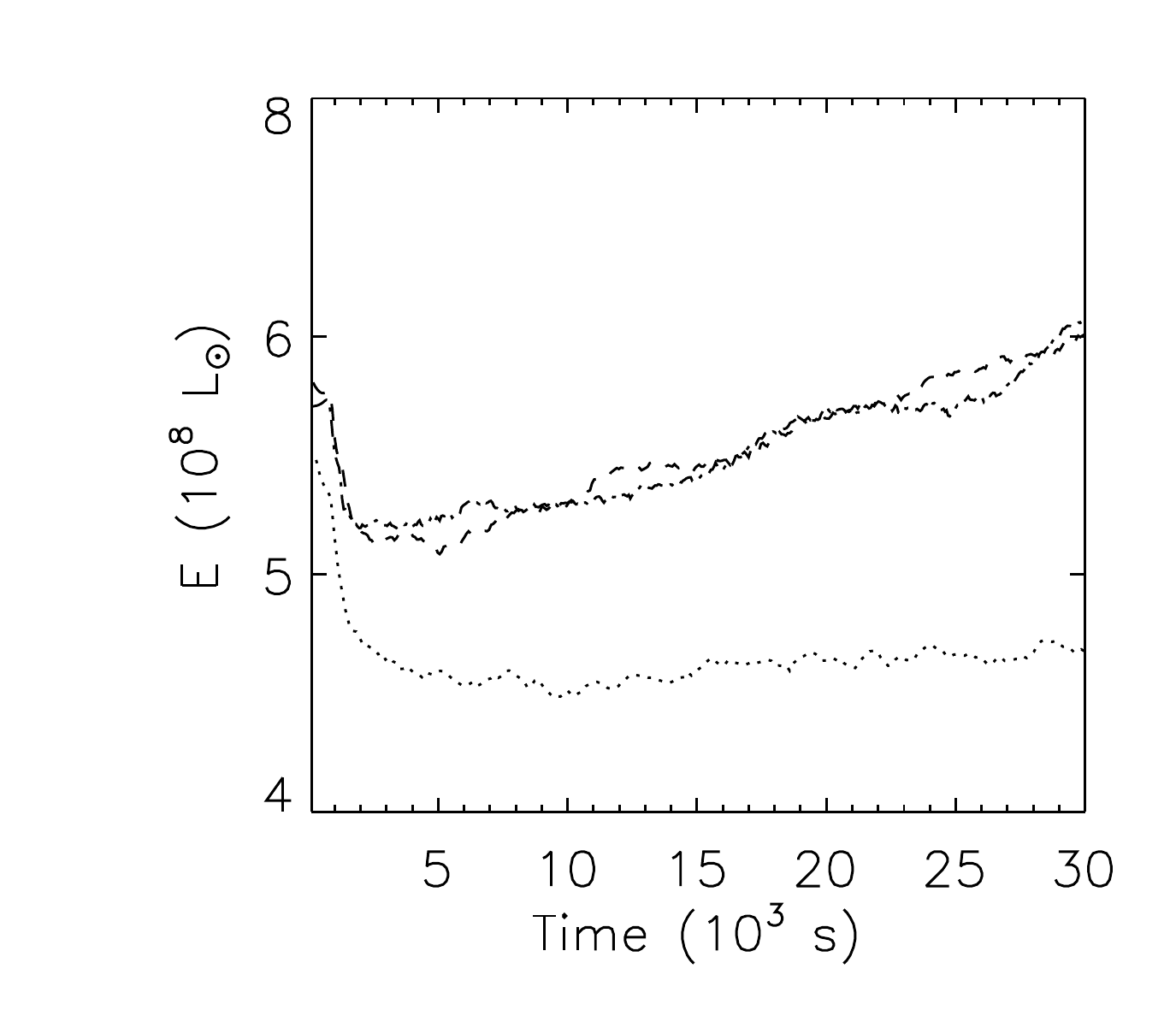}
\includegraphics[width=0.49\hsize]{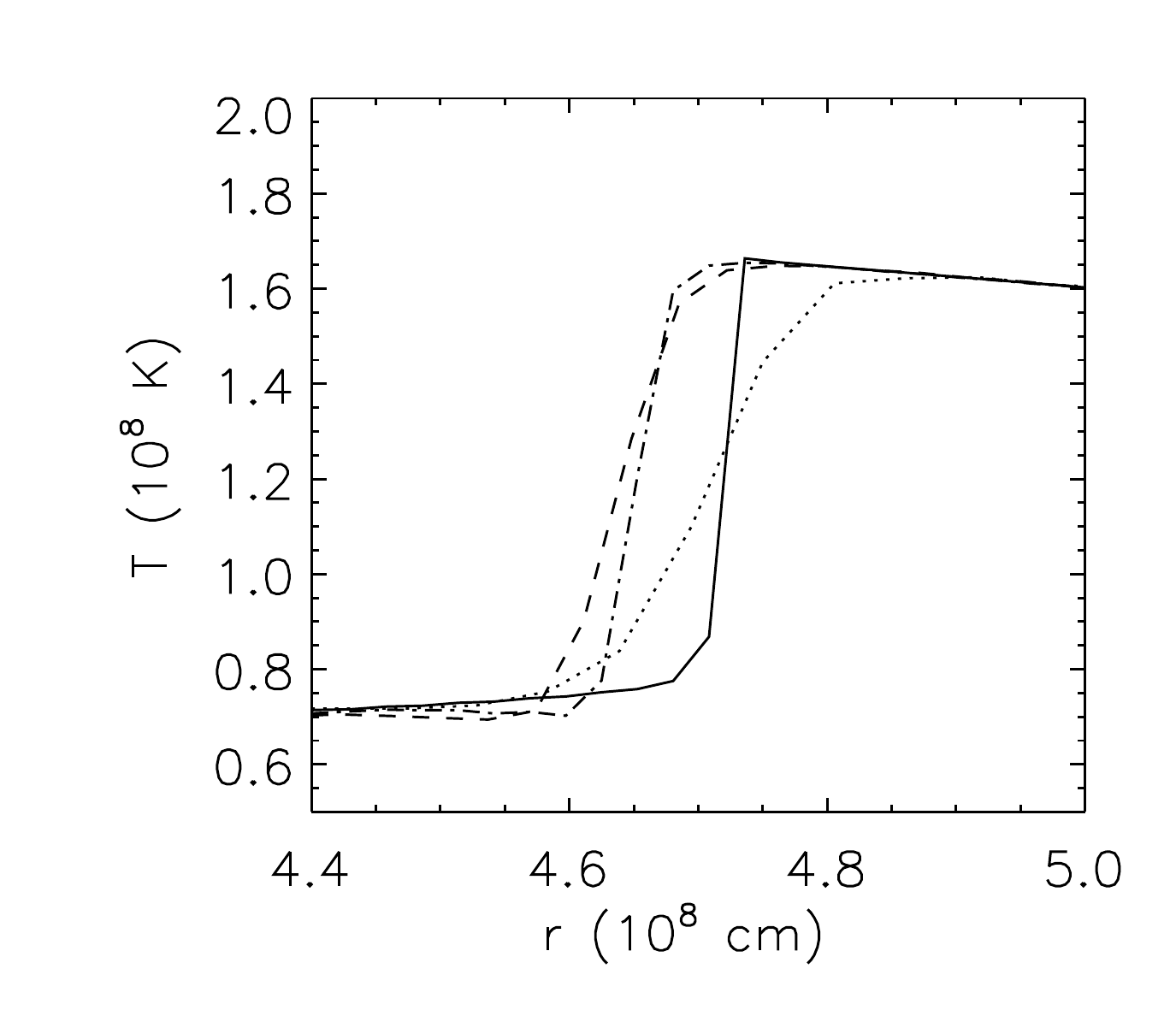}
\caption{{\it{Left panel:}} Evolution of the total energy production
         rate in solar luminosity L$_{\odot}$ for models DV2 (dotted),
         DV3 (dashed), and DV4 (dash-dotted), respectively.
         {\it{Right panel:}} Mean temperature distribution near the
         temperature inversion for models DV2 (dotted), DV3 (dashed),
         and DV4 (dash-dotted) at a $t = 30000\,$ s, respectively. The
         initial distribution is shown by the solid line.}
\label{fig5.8.9}  
\end{figure*} 

We have also found an influence of the turbulent entrainment on the
outer boundary of the convection zone. In the initial model this
boundary is located at $r = 9.2\, 10^{8}\,$cm and corresponds to a
discontinuous change in the distribution of elements
(Fig.\,\ref{fig2.2.3}), which in stellar evolution models results from
the assumed instantaneous mixing. In such models all species in the
convectively unstable region are mixed instantaneously across the
whole convection zone, while the regions which are assumed to be
convectively stable do not experience any mixing at all.

The distribution of $^{12}$C at the end of our simulation DV4 is
depicted in Fig.\,\ref{fig5.6.7}. Compared to the initial model there
is a clear shift of the carbon discontinuity at the outer edge of the
convection zone to a larger radius ($r = 9.7\, 10^{8}\,$cm). In
hydrodynamic simulations the gas overshoots naturally from the
convectively unstable to the formally convectively stable region
because of its inertia. At the boundaries of the convection zone the
overshooting seems always to destroy the stability according to the
Schwarzschild criterium transforming the originally convectively
stable region into a convectively unstable one. This allows the
boundary to propagate further when a subsequent load of gas will try
to overshoot at a later time. We have estimated the propagation speed
of the outer boundary of the convection zone to be about $\sim
14\,\mes$. With a propagation speed of this order the convection zone
would reach the hydrogen rich layers surrounding the helium core at a
radius $r = 1.9\, 10^{9}\,$cm and trigger a hydrogen injection flash
\citep{SchalttlCassisiSalaris2001} within just 10 days. Expected
hydrodynamic phenomena due to the extra hydrogen mixing into the
helium burning shell via such an extended convection zone could alter
the structure of the star significantly. Moreover, additional
nucleosynthesis could be triggered, since the hydrogen entrainment
will result in the production of neutrons and possibly also of some
s-process elements. The hydrogen injection flash in Pop\,I stars is in
contradiction to the canonical scenario since stellar evolutionary
models fail to inject hydrogen to the helium core during the core
helium flash, unless their metallicity is close to zero
\citep{FujimotoIben1990}.

Since the turbulent entrainment at the inner convective boundary
involved just three radial grid zones over the longest simulations we
performed, the estimated propagation velocity must be taken with care and be
considered as an order of magnitude estimate. The turbulent
entrainment at the outer convective boundary involved eighteen 
numerical zones in radial dimension, therefore the estimated
propagation velocity has higher confidence level, but still it should 
be taken as a rough number.
 
\subsubsection{Two-dimensional models with different resolution}
We find only minor differences between the properties of model DV4 and
those of the corresponding models computed with a different grid
resolution.

First, the initial mapping process leads to different interpolation
errors for different grid resolutions.  However, the major source of
discrepancy in this phase of the calculation is the stabilization
itself. The iterative procedure which minimizes the numerical fluxes
across zone boundaries (in order to keep the model in hydrostatic
equilibrium) tends to decrease the temperature stronger in models with
lower resolution.

Another source of discrepancy is caused by the numerical diffusion
which is obviously larger in models with lower resolution. Therefore,
model DV2 suffers more from numerical diffusion than model DV3 or DV4,
which is evident from Figure\,\ref{fig5.8.9}. The temperature inversion,
which is almost discontinuous at the beginning, gets smoothed out
faster in model DV2. Note, that the temperature inversion is situated
at smaller radii for models with higher resolution, since the typical
flow velocities are higher in better resolved models
(Tab.\,\ref{simu2d}), \ie the turbulent entrainment is more effective,
and the temperature inversion propagates with higher speed.

Nevertheless, models DV3 and DV4 seem to be well resolved since their
mutual differences are minor. The temporal evolution of their total
nuclear energy production rate, for instance, overlaps almost
perfectly (Fig.\,\ref{fig5.8.9}).  The temperature fluctuations in the
two-dimensional models are suppressed stronger in the better resolved
models. Contrary to \citet{Dearborn2006}, the more intense temperature
fluctuations occurring in models that we have calculated with grid
resolutions even lower than that of model DV2, did not lead to an
explosion.

\section{Summary}
We have presented one and two-dimensional (\ie axisymmetric)
hydrodynamic simulations of the core helium flash near its peak
covering about eight hours of evolution time.  We find no hydrodynamic
events which deviate significantly from the prediction of stellar
evolutionary calculations. After an initial adjustment phase the 2D
models reach a quasi-steady state where the temperature and nuclear
energy production rate are only slowly increasing.

Convection plays a crucial role in keeping the star in hydrostatic
equilibrium. Based on our two-dimensional simulation with the highest
grid resolution (model DV4), convection follows approximately the
predictions of mixing length theory, although the temperature gradient
of our dynamically evolved 2D models deviates by about 1\% from that
of the initial model which is obtained from (1D) stellar evolutionary
calculations.  The maximum temperature $\langle T \rangle_{max}$ in
out best resolved model DV4 rises with a rate of about $40\,\Ks$,
which is about 60\% smaller than the rate predicted by stellar
evolutionary calculations. The mean convective velocity exceeds the
velocities predicted by mixing length theory by up to factor of four.

During the early 2D dynamic evolution the size of the convective
region does not deviate from that of the initial (hydrostatic)
model. However, after a stable convective pattern is established, our
2D simulations show that the convective flow, consisting of four
quasi-stationary large scale ($\sim 40\,$ degrees angular width)
vortices, starts to push the inner and the outer boundary of the
convection zone as determined by the Schwarzschild stability criterium
towards the center of the star, and towards the stellar surface,
respectively. This results in a rapid growth of the radial extent of
the convection zone on dynamic timescales.

Our 2D simulations further suggest that it is unlikely that the core
helium flash is followed by subsequent core helium mini-flashes, which
are observed in (1D) stellar evolutionary calculations, since the
inner convective boundary could reach the center of the core in less
than one month.  On the other hand, the injection of hydrogen from the
stellar envelope into the helium core is likely to happen within just
10 days, which is in contradiction to the predictions of the canonical
evolution of low-mass Pop\,I stars.

As our 2D axisymmetric simulations probably cannot properly capture
the intrinsically three-dimensional character of the convective flow,
we have started to perform also 3D simulations of the core helium
flash. In addition, we plan to extend our 2D simulations to time
intervals of several days instead of hours. The results of these
long-term 2D simulations and of the first well resolved 3D simulations
of the core helium flash will be presented in due time elsewhere.

\begin{acknowledgements}
The calculations were performed at the Rechenzentrum Garching on the
IBM pSeries Power5 system, and at the Leibniz-Rechenzentrum of the
Bavarian Academy of Sciences and Humanities on the SGI Altix 4700
system.  The authors want to thank Frank Timmes for some of his public
Fortran subroutines which we used in the Herakles code for calculating
the core helium flash models. We also thank Kurt Achatz, whose
unpublished hydrodynamic simulations of the core helium flash,
performed as part of his diploma work, have motivated and inspired us.
\end{acknowledgements}

\appendix 
 
\section{Energy fluxes} 
An analysis of the vertical energy transport allows for conclusions
about the importance of the different physical processes occurring in
the convection zone. To separate the various contributions to the
total energy flux \citep{HurlburtToomre1986,Achatz1995}, one 
integrates the hydrodynamic equation of energy conservation
\begin{equation}
\begin{array}{rr} 
   \partial_t(\rho e) + \partial_i(v_i(\rho e+p) - v_j\Sigma_{ij} 
         - K \partial_iT) = -\rho v_i\partial_i\Phi \quad,\quad &  \\ 
     i,j=1,2,3 &  
\end{array}
\end{equation} 
(with $e=\varepsilon+v_iv_i/2$ being the specific total energy
density) over angular coordinates ($\theta$, $\phi$), and separates
both the specific enthalpy ($\varepsilon + p/\rho$) and the kinetic
energy ($v_iv_i/2$) into a horizontal mean and a perturbation
($f \equiv \overline f+f'$). This results in
\begin{equation} 
   \partial_tE + \partial_r(F_C+F_K+F_R+F_V+F_E)=0 
\label{aformequenE} 
\end{equation} 
with 
\footnote{The gravitational potential $\Phi$ was assumed to be constant
          for simplicity.}
\newcommand{\dOm}{\,r^2\ad\Omega} 
\begin{eqnarray} 
   E   &=& \oint \rho e \dOm \\ 
   F_C &=& \oint v_r\rho\cdot \left(\varepsilon+\frac{p}{\rho}\right)' 
           \dOm \\ 
   F_K &=& \oint v_r\rho\cdot \left(\frac12 v_iv_i\right)' \dOm 
           \quad,\quad i=1,2,3\\ 
   F_R &=& -\oint K\partial_rT \dOm \\ 
   F_V &=& -\oint v_i \Sigma_{ri} \dOm \quad,\quad i=1,2,3\\ 
   F_E &=& 4\pi r^2\overline{v_r\rho}\cdot 
           \left(\,\overline{\varepsilon+\frac p\rho}+\overline{\frac12v_iv_i} 
            +\partial_r\Phi\,\right) \,. 
\label{app_fe}
\end{eqnarray} 
Here, the various terms $F_i$ give the total energy transported per
unit time across a sphere by different physical processes. They are
the convective (or enthalpy) flux, $F_C$, the flux of kinetic energy,
$F_K$, the flux by heat conduction and radiation, $F_R$, and the
viscous flux, $F_V$. Finally, $F_E$, collects all terms causing a
spherical mass flow, \ie the model's expansion or contraction, while
$F_C$ and $F_K$ rest on deviations from this mean energy flow
(vortices). The latter are the major contributors to the heat
transport by convection, while $F_V$ is usually negligibly small.

In a similar way one can also formulate a conservation equation for
the mean horizontal kinetic energy, which provides further insight
into the effects of convective motions. Using the other hydrodynamic
equations
\begin{eqnarray}
   \partial_t(\rho)+\partial_i(\rho v_i) &=& 0 \\ 
   \partial_t(\rho v_i)+\partial_j(\delta_{ij}p+\rho v_iv_j-\Sigma_{ji}) 
      &=& -\rho\partial_i\Phi \quad,\quad \\
   &  & i,j=1,2,3 
\end{eqnarray} 
and $\partial_t(\rho v_iv_i/2) = v_i \partial_t(\rho v_i) - v_i v_i
\partial_t\rho/2$, one finds
\begin{equation} 
   \partial_tE_K + \partial_r(F_K+F_P+F_V+F_{E,K}) 
        = P_A + P_P + P_V + P_{E,K} 
      \label{aformequenEK} 
\end{equation} 
With $F_K$ and $F_V$ as introduced above, one obtains
\begin{eqnarray} 
   E_K &=& \oint \frac\rho2 v_i v_i \dOm \\ 
   F_P &=& -\oint v_r p' \dOm \\ 
   F_{E,K} &=& 4\pi r^2\overline{v_r\rho}\cdot 
              \left(\,\overline{\frac p\rho+\frac{v_i v_i}2}\,\right) \\ 
   P_A &=& -\oint v_r\rho'\partial_r\Phi \dOm \\ 
   P_P &=& \oint p'\partial_i v_i \dOm \\ 
   P_V &=& -\oint \partial_iv_j\cdot\Sigma_{ij} \dOm \\ 
   P_{E,K} &=& 4\pi r^2\cdot 
          \left(\,\overline p\,\overline{\partial_i v_i}- 
           \overline v_r\overline\rho\,\partial_r\Phi\,\right) 
    \quad,\quad i=1,2,3\quad 
\end{eqnarray} 
where the $P_i$ are source or sink terms of the kinetic energy. They
are separated into the effect of buoyancy forces ($P_A$), friction
forces ($P_V$), and the work due to density fluctuations ($P_P$,
volume changes). By analyzing the various $P_i$ one can determine what
brakes or accelerates convective motions. The acoustic flux, $F_P$,
describes the vertical transport of density fluctuations. $F_{E,K}$
and $P_{E,K}$ describe the effect of expansion (volume work, and work
against the gravitational potential), similar to $F_E$ in
Eq.\,(\ref{app_fe}).

\bibliography{referenc}

\end{document}